\documentclass[aip,groupedaddress,reprint]{revtex4-1}
\usepackage{graphicx}
\usepackage{amsmath}
\usepackage{hyperref}
\draft % marks overfull lines with a black rule on the right

\begin{document}

\title{Spin-accumulation capacitance and its application to magnetoimpedance}

\author{Yao-Hui Zhu}
\email[Electronic mail:]{yaohuizhu@gmail.com}
\author{Xiao-Xue Zhang}
\author{Jian Liu}
\author{Pei-Song He}
\affiliation{Physics Department, Beijing Technology and Business University, Beijing 100048, China}

\date{\today}

\begin{abstract}
It has been known that spin-dependent capacitances usually coexist with geometric capacitances in a magnetic multilayer. However, the charge and energy storage of the capacitance due to spin accumulation has not been fully understood. Here, we resolve this problem starting from the charge storage in the spin degree of freedom: spin accumulation manifests itself as an excess of electrons in one spin channel and an equal deficiency in the other under the quasi-neutrality condition. This enables us to model the two spin channels as the two plates of a capacitor. Taking a ferromagnet/nonmagnet junction as an example and using a method similar to that for treating quantum capacitance, we find that a spin-accumulation (SA) capacitance can be introduced for each layer to measure its ability to store spins. A spatial charge storage is not essential for the SA capacitor and the energy stored in it is the splitting energy of the spin-dependent chemical potentials instead of the electrostatic energy. The SA capacitance is essentially a quantum capacitance due to spin accumulation on the scale of the spin-diffusion length. The SA capacitances can be used to reinterpret the imaginary part of the low-frequency magnetoimpedance.
\end{abstract}

%\pacs{}% insert suggested PACS numbers in braces on next line

\maketitle %\maketitle must follow title, authors, abstract and \pacs

\section{Introduction}

In many applications, capacitances have a remarkable influence on the speed and power dissipation of the devices, and also set an upper frequency limit (cutoff frequency) for their correct operations.~\cite{Neamen2012,Theis2010} The effects of the capacitances on spintronic devices have also been observed and analyzed recently. The magnetocapacitance or magnetoimpedance has been studied, for example, in magnetic tunnel junctions (MTJs)~\cite{Xiao01,Chui02,Kaiju2002,Chien2006,Gupta07,Chang2010,Kaiju2015,Parui2016} and in a single electron transistor.~\cite{Lee15}

% Object: open questions

Using a time-dependent approach, Rashba has studied the frequency-dependent impedance of a junction composed of a ferromagnetic (FM) conductor, a spin-selective tunnel or Schottky contact, and a nonmagnetic (NM) conductor.~\cite{Rashba02a} The imaginary part of the impedance was attributed to a diffusion capacitance $C_\mathrm{diff}$, which can be compared to (but is certainly different from) that of a p-n junction.~\cite{Neamen2012} However, Rashba's derivation involved the quasi-neutrality condition, which assumes that the charge accumulation is negligible everywhere. Thus it seems difficult to reconcile this condition with the charge storage that is usually associated with the diffusion capacitance.~\cite{Neamen2012} In the present paper, we try to resolve this problem by introducing a \emph{spin-accumulation} (SA) capacitance for each layer. The two plates of the SA capacitor model the two spin channels, since spin accumulation manifests itself as an excess of electrons in one spin channel and an equal deficiency in the other under the quasi-neutrality condition. Thus the charge storage of the SA capacitor happens in the spin degree of freedom rather than the coordinate space, and a spatial charge storage is not essential to it. We also prove that the SA capacitances lead to the same low-frequency reactance as $C_\mathrm{diff}$. Then the conflict mentioned above is resolved.

% This new type of capacitance measures the ability of a material to store spins instead of charge.

%although it is present together with the spin accumulation in the FM layers and around the interface

The form of energy storage associated with the diffusion capacitance is another basic question, because it indicates the type and origin of the capacitance. However, this issue has not been addressed to our knowledge.~\cite{Neamen2012,Rashba02a} On the other hand, most studies on the magnetoimpedance of the MTJs assumed that the energy is stored in an extra electrostatic field like usual geometric capacitances.~\cite{Xiao01,Chien2006} Here, we also show that the energy storage associated with the magnetoimpedance of the FM/NM junction is equal to that stored in the SA capacitors. It is the splitting energy of the spin-dependent chemical potentials instead of the electrostatic energy. This characteristic is similar to that of a quantum capacitance, which stores energy in forms other than the electrostatic field.~\cite{Luryi88,Datta2005,Kopp2009}

We study the SA capacitance using a method similar to that for dealing with quantum capacitance.~\cite{Luryi88} This method is quite different from Rashba's time-dependent approach.~\cite{Rashba02a} It enables us to find the connections between the SA capacitance, the diffusion capacitance, and the quantum capacitance. Moreover, this method has the potential to be used in more complicated situations.

% We can write the SA capacitance in two forms: one is obviously similar to the diffusion capacitance of a p-n junction~\cite{Neamen2012} and the other is analogous to the quantum capacitance.~\cite{Luryi88} Thus the SA capacitance has the characteristics of the diffusion capacitance and the quantum capacitance.

% As an application, we rewrite the low-frequency impedance of the FM/NM junction in terms of the SA capacitances by using an equivalent circuit.

% The SA capacitance gives rise to an electro-motive force when `discharging' in lateral spin valves.

% There are still controversies over magnetocapacitance effect from both theoretical and experimental aspects~\cite{Gupta07,Ingvarsson10,Yang12}.

This paper is organized as follows. In section~\ref{theory}, we introduce the SA capacitance for the FM/NM junction. Then the model is used to reinterpret the magnetoimpedance in section~\ref{discussion}. The main conclusions are given in section~\ref{conclusion}.

\section{Spin-accumulation capacitance\label{theory}}

\begin{figure}
\includegraphics[width=0.48\textwidth]{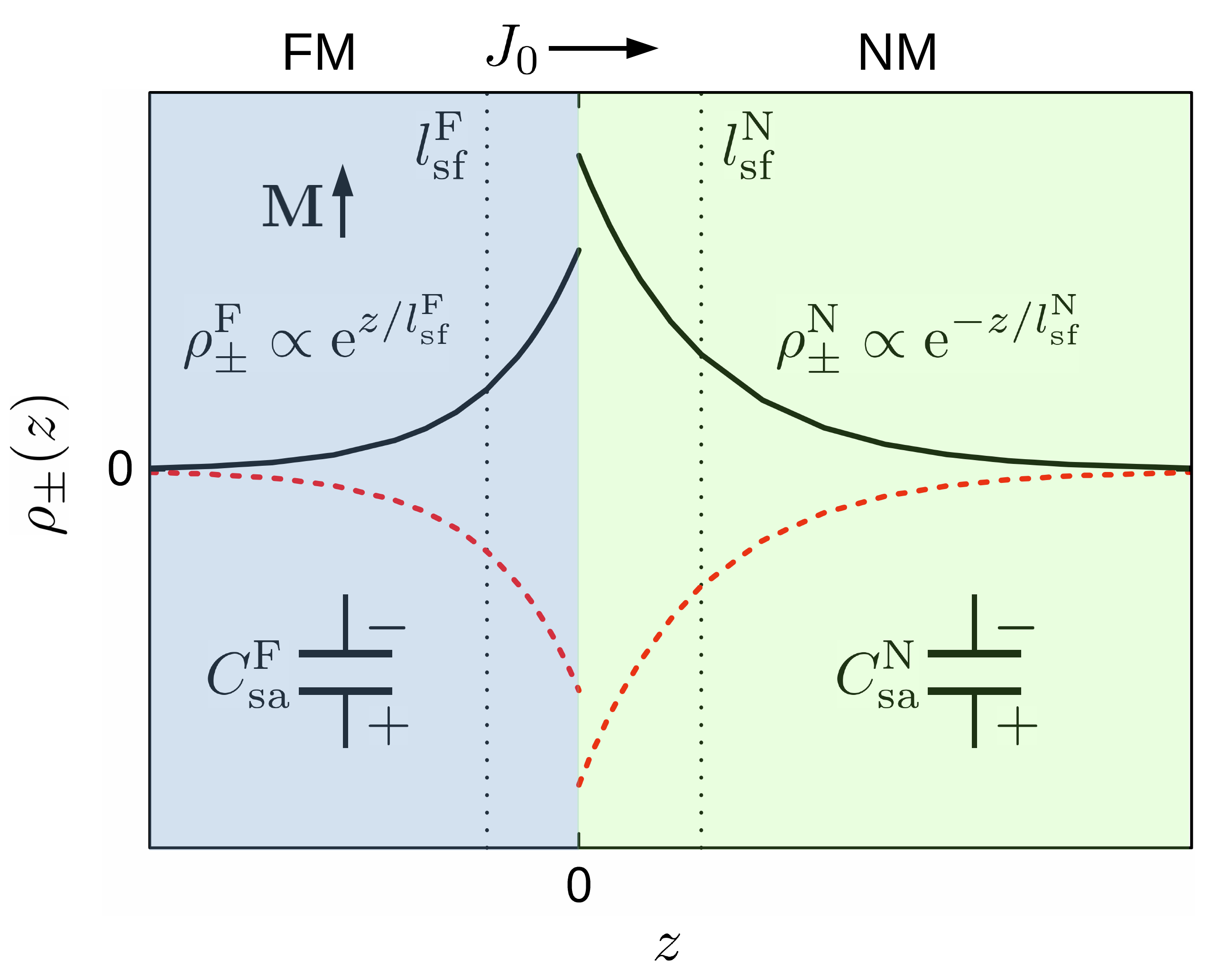}
\caption{\label{fig_fin} Sketch of the spin-dependent charge density $\rho_\pm(z)$ in the FM/NM junction. We plot the curves using Eq.~(\ref{rhomu}) but without realistic parameters, because a demonstration of the basic profile is enough for our current purpose. The units for $\rho_\pm(z)$ and $z$ are arbitrary. The constant DC current has a density of $J_0$. The solid curves on the left and right sides of the interface at $z=0$ stand for $\rho_+(z)$ in the FM and NM layers, respectively. Similarly, the dashed curves are for $\rho_-(z)$. The vertical dotted lines indicate the scale of the spin-diffusion lengths, $l_\mathrm{sf}^\mathrm{F}$ and $l_\mathrm{sf}^\mathrm{N}$, in the FM and NM layers, respectively. The capacitances $C_\mathrm{sa}^\mathrm{F}$ and $C_\mathrm{sa}^\mathrm{N}$ are defined by Eq.~(\ref{sc}), and will be connected in the circuit shown by Fig.~\ref{fig_circuit_fin}(b).}
\end{figure}

To illustrate the concept of the SA capacitance, we consider a magnetic multilayer with current perpendicular to the plane, which is the well-known configuration giving rise to spin accumulation.~\cite{vf93} To be specific, we consider the same FM/NM junction studied by Rashba.~\cite{Rashba02a} The junction is composed of a ferromagnetic layer occupying $z<0$, a spin-selective contact at $z=0$, and a nonmagnetic layer occupying $z>0$. The $z$ axis is set to be perpendicular to the layer plane as shown in Fig.~\ref{fig_fin}. Without loss of generality, the magnetization of the FM is set to be ``up'' and the current is flowing in the positive direction of the $z$ axis.

Although the importance of capacitance usually shows up in time-dependent transport, it is also present in the constant DC situation, which is much easier to deal with. The spin accumulation is usually described by the splitting of the spin-dependent chemical potentials
\begin{equation}\label{spinaccu}
\mu_\mathrm{m}(z)=2\Delta\mu(z)=\mu_{+}(z)-\mu_{-}(z),
\end{equation}
where the subscripts ``$+$'' and ``$-$'' stand for the absolute spin directions ``up'' and ``down'', respectively.~\cite{vf93} In each layer, $\Delta\mu(z)$ satisfies the well-established spin diffusion equation
\begin{equation}
\frac{\partial^2\Delta\mu(z)}{\partial{z}^2}=
\frac{\Delta\mu(z)}{l_\mathrm{sf}^2},\label{new6a}
\end{equation}
where $l_\mathrm{sf}$ is the spin-diffusion length (see Appendix~\ref{appfn}). The solution to Eq.~(\ref{new6a}) in each layer is of exponential form as shown by Eqs.~(\ref{deltamuf}) and (\ref{deltamun}). Introducing the average chemical potential $\mu(z)=[\mu_{+}(z)+\mu_{-}(z)]/2$ and using Eq.~(\ref{spinaccu}), we can write the spin-dependent chemical potentials as
\begin{equation}
\mu_\pm(z)=\mu(z)\pm\Delta\mu(z).\label{averagemu}
\end{equation}
Since we are concerned with capacitance, it is more convenient to use the charge density in the two spin channels
\begin{equation}
\rho_\pm(z)=-eN_s\left[\mu_\pm(z)-\mu^0\right],\label{deltans}
\end{equation}
where $-e$ is the charge of an electron and $N_s$ the density of states at the Fermi level $\mu^0$.~\cite{zhu08,zhu14} According to the Valet-Fert theory,~\cite{vf93} $N_s$ is assumed to be the same for the two spin directions and also for both layers. We have $\mu(z)=\mu^0$ in the NM layer, whereas $\mu(z)$ is unequal to $\mu^0$ in general for the FM layer.~\cite{zhu14} However, it can be justified for our problem to assume
\begin{equation}
\mu(z)=\mu^0\label{qnd1}
\end{equation}
even in the FM layer using the quasi-neutrality approximation (see Appendix~\ref{qnd}). Then substituting Eqs.~(\ref{averagemu}) and (\ref{qnd1}) into Eq.~(\ref{deltans}), we have
\begin{equation}\label{rhomu}
\rho_\pm(z)=\mp{e}N_s\Delta\mu(z),
\end{equation}
which is valid in both FM and NM layers. Therefore, the electron excess in one spin channel cancels the deficiency in the other, $\rho_{+}(z)+\rho_{-}(z)=0$, in both layers.

%https://en.wikipedia.org/wiki/Diffusion_capacitance

%The spin accumulation $\Delta\mu(z)$ can be determined by the well-established method~\cite{vf93} and results are outlined in \ref{appfn}.

The basic idea of the SA capacitance arises initially from the resemblance of charge storage between the two spin channels and the two plates of a normal capacitor, as shown in Fig.~\ref{fig_fin}. Further analysis shows that the spin accumulation also bears similarity to a capacitor in other aspects, including energy storage, leakage current, and heat generation.

%\paragraph{Charge storage}
\subsection{Charge storage}

When the two spin channels are modeled as the two plates of the SA capacitor, the charge storage of the capacitor can be defined as the absolute value of the charge accumulation in either spin channel. Treating the FM and NM layers separately, we can write the charge storage as
\begin{align}\label{chargefn}
Q_\mathrm{sa}^\mathrm{F}&=\int_{-\infty}^0\left|\rho_\pm(z)\right|\mathrm{d}z
=\int_{-\infty}^0{e}N_s\left|\Delta\mu(z)\right|\mathrm{d}z,\\
Q_\mathrm{sa}^\mathrm{N}&=\int_0^\infty\left|\rho_\pm(z)\right|\mathrm{d}z
=\int_0^\infty{e}N_s\left|\Delta\mu(z)\right|\mathrm{d}z,
\end{align}
where Eq.~(\ref{rhomu}) has been used. Following the procedure detailed in Appendix~\ref{appfn}, we can write $Q_\mathrm{sa}^\mathrm{F}$ and $Q_\mathrm{sa}^\mathrm{N}$ in a form resembling a normal capacitor
\begin{equation}\label{qs}
Q_\mathrm{sa}^\mathrm{F}=C_\mathrm{sa}^\mathrm{F}V_\mathrm{c}^\mathrm{F},\qquad
Q_\mathrm{sa}^\mathrm{N}=C_\mathrm{sa}^\mathrm{N}V_\mathrm{c}^\mathrm{N},
\end{equation}
where we have introduced the SA capacitances
\begin{equation}\label{sc}
C_\mathrm{sa}^\mathrm{F}=\frac{T_1^\mathrm{F}}{2r_\mathrm{F}},\qquad
C_\mathrm{sa}^\mathrm{N}=\frac{T_1^\mathrm{N}}{2r_\mathrm{N}},
\end{equation}
and the corresponding effective voltages
\begin{equation}\label{ep}
V_\mathrm{c}^\mathrm{F}=\frac{\left|\Delta\mu(0^-)\right|}{e},\qquad
V_\mathrm{c}^\mathrm{N}=\frac{\left|\Delta\mu(0^+)\right|}{e},
\end{equation}
for the \emph{semi-infinite} FM and NM layers, respectively. In the expression of $C_\mathrm{sa}^\mathrm{F(N)}$ given in Eq.~(\ref{sc}), $T_1^\mathrm{F}$ ($T_1^\mathrm{N}$) is the spin-relaxation time in the FM (NM) layer.~\cite{zhu08} The resistances $r_\mathrm{F}$ and $r_\mathrm{N}$ are defined as~\cite{fert01}
\begin{equation}\label{rfn}
r_\mathrm{F}=\rho^\ast_\mathrm{F}l_{\mathrm{sf}}^\mathrm{F},\qquad
{r}_\mathrm{N}=\rho^\ast_\mathrm{N}l_{\mathrm{sf}}^\mathrm{N},
\end{equation}
where $\rho^\ast_\mathrm{F}$ ($\rho^\ast_\mathrm{N}$) is the FM (NM) resistivity. Using the expressions of $\Delta\mu(0^-)$ and $\Delta\mu(0^+)$ in Eq.~(\ref{mu0pm}), we can rewrite Eq.~(\ref{ep}) in a form similar to Ohm's law
\begin{equation}\label{ep2}
V_\mathrm{c}^\mathrm{F}=r^\mathrm{sf}_\mathrm{F}\left|\alpha_\mathrm{F}\right|\frac{J_0}{2},\qquad
V_\mathrm{c}^\mathrm{N}=r^\mathrm{sf}_\mathrm{N}\left|\alpha_\mathrm{N}\right|\frac{J_0}{2},
\end{equation}
where we have defined the dimensionless parameters
\begin{equation}\label{alphafn}
\alpha_\mathrm{F}=\frac{\beta(r_\mathrm{N}+r_\mathrm{b}^\ast)-\gamma{r}_\mathrm{b}^\ast}
{r_\mathrm{F}+r_\mathrm{b}^\ast+r_\mathrm{N}},\qquad
\alpha_\mathrm{N}=\frac{\beta{r}_\mathrm{F}+\gamma{r}_\mathrm{b}^\ast}
{r_\mathrm{F}+r_\mathrm{b}^\ast+r_\mathrm{N}},
\end{equation}
and the characteristic (spin-flip) resistances
\begin{equation}\label{rsr}
r^\mathrm{sf}_\mathrm{F}=2r_\mathrm{F},\qquad{r}^\mathrm{sf}_\mathrm{N}=2r_\mathrm{N},
\end{equation}
for the FM and NM layers, respectively. In Eq.~(\ref{alphafn}), the bulk spin asymmetry coefficient $\beta$ in the FM layer is defined by the relation
\begin{equation}
1/\sigma_{\uparrow(\downarrow)}=2\rho_\mathrm{F}^\ast[1-(+)\beta],
\end{equation}
where $\sigma_{\uparrow(\downarrow)}$ is the conductivity for the majority (minority) spin direction. Similarly, we have $1/\sigma_{\pm}=2\rho_\mathrm{N}^\ast$ in the NM layer. The interfacial resistance $r_\mathrm{b}^\ast$ and its spin asymmetry coefficient $\gamma$ are defined by
\begin{equation}\label{bcr}
r_{\uparrow\left(\downarrow\right)}=2r_{b}^{*}\left[1-\left(+\right)\gamma\right],
\end{equation}
where $r_{\uparrow\left(\downarrow\right)}$ is the resistance of the majority (minority) spin channel. In general, the effective voltages $V_\mathrm{c}^\mathrm{F}$ and $V_\mathrm{c}^\mathrm{N}$ are not equal to each other due to the interface resistance. In the configuration specified above (``up'' magnetization and positive DC current), $\alpha_\mathrm{N}$ is always positive, whereas $\alpha_\mathrm{F}$ may be negative. Their interpretation will be given by Eq.~(\ref{jleak}) in combination with the effective circuit shown in Fig.~\ref{fig_circuit_fin}(b).

The expression of $C_\mathrm{sa}^\mathrm{F(N)}$ in Eq.~(\ref{sc}) is similar to that of the diffusion capacitance in a p-n junction.~\cite{Neamen2012} However, $C_\mathrm{sa}^\mathrm{F(N)}$ does not require a spatial charge storage, which is essential for the diffusion capacitance in a p-n junction. Therefore, $C_\mathrm{sa}^\mathrm{F(N)}$ measures the ability of a material to store spins instead of charge. Spin accumulating and dissipating correspond to charging and discharging of the SA capacitor, respectively. Note that $C_\mathrm{sa}^\mathrm{F(N)}$ cannot be compared directly with the diffusion capacitance $C_\mathrm{diff}$ in Eq.~(\ref{cdiff}), which was introduced by Rashba.~\cite{Rashba02a} One distinction between them is the independence of $C_\mathrm{sa}^\mathrm{F(N)}$ on the spin asymmetry coefficient $\beta$ or $\gamma$.

\subsection{Energy storage}

Since $\mu_\mathrm{m}(z)$ is the splitting between the chemical potentials of the two spin channels, spin accumulation also accompanies an increase in energy in comparison to the equilibrium state.~\cite{Tulapurkar11,Juarez16} The energy stored in the differential $\mathrm{d}z$ is equivalent to the energy required to shift a number of electrons, $N_s|\Delta\mu(z)|\mathrm{d}z$, from the spin channel with lower chemical potential to the other. Because the average energy increase per electron is just $|\Delta\mu(z)|$, the energy storage in $\mathrm{d}z$ is $N_s[\Delta\mu(z)]^2\mathrm{d}z$. The total energy storage is the sum of the following two terms
\begin{align}
W_\mathrm{sa}^\mathrm{F}=\int_{-\infty}^0{N}_s\left[\Delta\mu(z)\right]^2\mathrm{d}z,\\
W_\mathrm{sa}^\mathrm{N}=\int_0^\infty{N}_s\left[\Delta\mu(z)\right]^2\mathrm{d}z,
\end{align}
where $W_\mathrm{sa}^\mathrm{F}$ and $W_\mathrm{sa}^\mathrm{F}$ are the contributions from the FM and NM layers, respectively. Using Eqs.~(\ref{deltamuf}) and (\ref{deltamun}), we can write them in a form resembling the energy stored in a capacitor
\begin{equation}\label{esc}
W_\mathrm{sa}^\mathrm{F}=\frac{1}{2}C_\mathrm{sa}^\mathrm{F}\left(V_\mathrm{c}^\mathrm{F}\right)^2,\qquad
W_\mathrm{sa}^\mathrm{N}=\frac{1}{2}C_\mathrm{sa}^\mathrm{N}\left(V_\mathrm{c}^\mathrm{N}\right)^2,
\end{equation}
where Eqs.~(\ref{sc}) and (\ref{ep}) have also been used.

The energy stored in an SA capacitor is essentially the splitting energy of the spin-dependent chemical potentials instead of the electrostatic energy associated with a geometric capacitance. This character is similar to the quantum capacitance, which stores energy in the form of the Fermi degeneracy energy.~\cite{Luryi88} In general, quantum capacitance appears when the spatial charge accumulation on at least one plate of the capacitor induces a change in chemical potential. This is also true for the SA capacitance according to the discussion above. Moreover, substituting Eq.~(\ref{csa}) into Eq.~(\ref{sc}), we can also rewrite the SA capacitance as
\begin{equation}\label{qc}
C_\mathrm{sa}^\mathrm{F(N)}=C_\mathrm{Q}l_\mathrm{sf}^\mathrm{F(N)},
\end{equation}
where we define $C_\mathrm{Q}$ as the quantum capacitance per unit volume and spin channel
\begin{equation}
C_\mathrm{Q}=e^2N_s,
\end{equation}
following Ref.~\onlinecite{Luryi88}. Thus we can also interpret the SA capacitance as the quantum capacitance due to spin accumulation on the scale of the spin-diffusion length. It has the characteristics of the diffusion capacitance as well as the quantum capacitance as shown by Eqs.~(\ref{sc}) and (\ref{qc}). However, the SA capacitance is different from the usual quantum capacitance: its change in chemical potential happens in the spin degree of freedom instead of the coordinate space owing to the relation $\mu(z)=\mu^0$.

%in addition to a variation in electrostatic potential
% Using Eq.~(\ref{ep}), one can also write $W_\mathrm{s}^\mathrm{F}$ and
% $W_\mathrm{s}^\mathrm{N}$ as
% \begin{equation}\label{esc}
% W_\mathrm{s}^\mathrm{F}=\frac{1}{2}C_\mathrm{s}^\mathrm{F}(r_\mathrm{F}\alpha_\mathrm{F})^2J_0^2,
% \qquad
% W_\mathrm{s}^\mathrm{N}=\frac{1}{2}C_\mathrm{s}^\mathrm{N}(r_\mathrm{N}\alpha_\mathrm{N})^2J_0^2.
% \end{equation}

\subsection{Leakage current}

\begin{figure}
\includegraphics[width=0.48\textwidth]{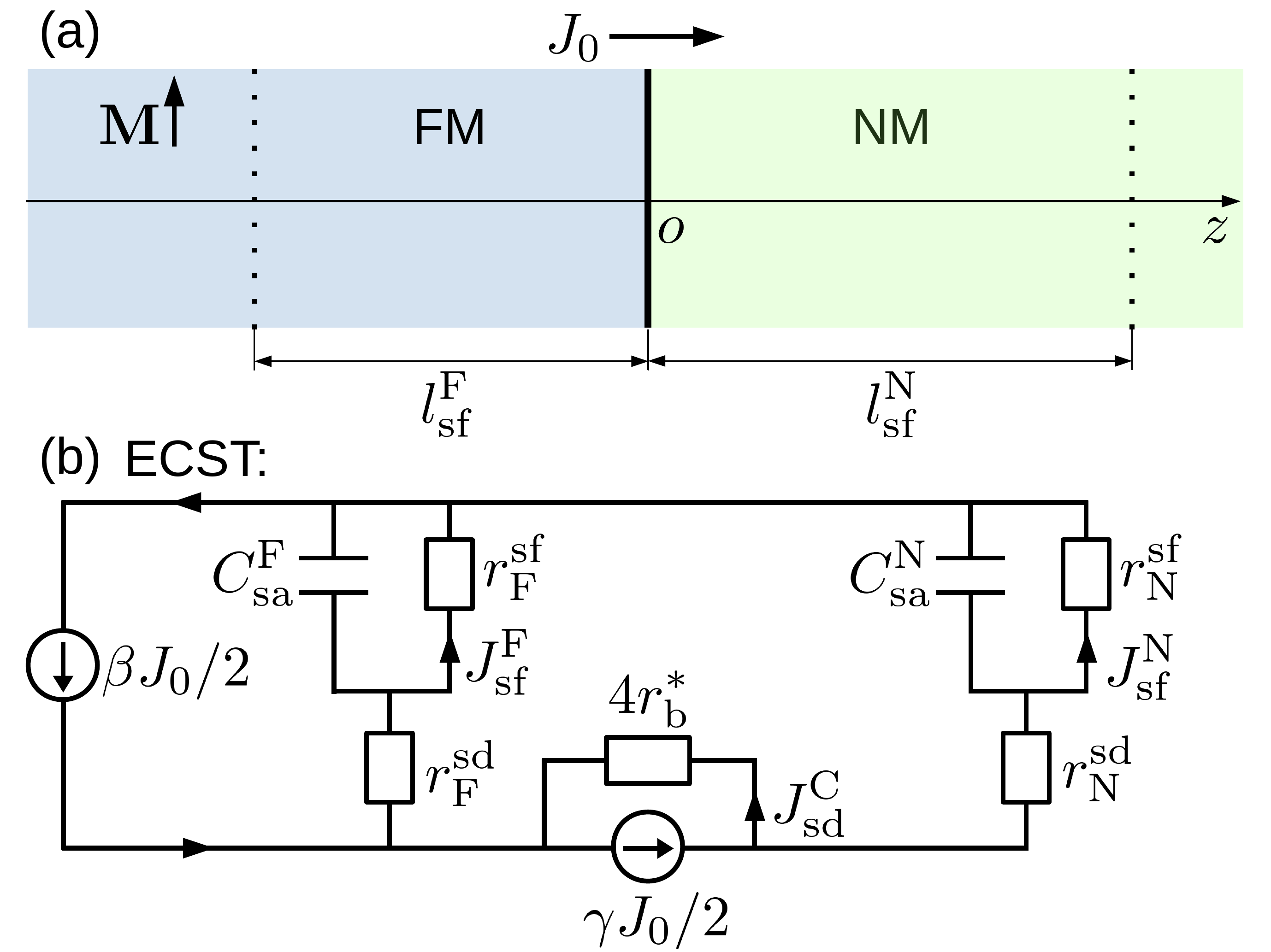}
\caption{\label{fig_circuit_fin}(a) Sketch of the FM/NM junction with the spin-selective contact, which is signified by the thick vertical line at $z=0$. The vertical dotted lines indicate the scales of spin-diffusion length in the two layers. (b) An effective circuit of the spin transport (ECST) in the junction shown by (a). The SA capacitors in Fig.~\ref{fig_fin} are connected in parallel with two characteristic resistors to model the leakage due to spin relaxation. The resistances labeled by $r^\mathrm{sf}_\mathrm{F}$ ($r^\mathrm{sf}_\mathrm{N}$) and $r^\mathrm{sd}_\mathrm{F}$ ($r^\mathrm{sd}_\mathrm{N}$) have the same value $2r_\mathrm{F}$ ($2r_\mathrm{N}$). The superscripts ``sf'' and ``sd'' stand for spin flip and spin diffusion [see the text right before Eq.~(\ref{rsd})], respectively. The current densities $J_\mathrm{sf}^\mathrm{F}$, $J_\mathrm{sf}^\mathrm{N}$, and $J_\mathrm{sd}^\mathrm{C}$ are given by Eqs.~(\ref{jleak}) and (\ref{jc}), respectively. The current densities $\beta{J}_0/2$ and $\gamma{J}_0/2$ are generated by two current sources, the FM layer [see Eq.~(\ref{leakcurrent})] and the spin-selective contact [see Eq.~(\ref{efc})], respectively.}
\end{figure}

Spin accumulation coexists with spin relaxation, in which electrons undergo transitions from the spin channel with higher chemical potential to the other via spin-flip scattering. This process can be modeled as the leakage of the SA capacitor (spin flux in Ref.~\onlinecite{Wegrowe2011}). If we set the positive direction of leakage current to be from the spin-down channel to the spin-up channel, the leakage current can be written as
\begin{align}
J_\mathrm{sf}^\mathrm{F}=\int_{-\infty}^0\frac{eN_{s}\mu_\mathrm{m}(z)}
{\tau_\mathrm{sf}^\mathrm{F}}\mathrm{d}z,\label{jsff}\\
J_\mathrm{sf}^\mathrm{N}=\int_0^\infty\frac{eN_{s}\mu_\mathrm{m}(z)}
{\tau_\mathrm{sf}^\mathrm{N}}\mathrm{d}z,\label{jsfn}
\end{align}
for the FM and NM layers, respectively. Here $\tau_{\rm{sf}}=2T_{1}$ is the spin-flip scattering time.~\cite{fl96} Using Eqs.~(\ref{deltamuf}) and (\ref{deltamun}), we can write $J_\mathrm{sf}^\mathrm{F}$ and $J_\mathrm{sf}^\mathrm{N}$ in a form resembling Ohm's law
\begin{equation}\label{jleak}
J_\mathrm{sf}^\mathrm{F}=\frac{V_\mathrm{c}^\mathrm{F}}{r^\mathrm{sf}_\mathrm{F}}
=\alpha_\mathrm{F}\frac{J_0}{2},\qquad
J_\mathrm{sf}^\mathrm{N}=\frac{V_\mathrm{c}^\mathrm{N}}{r^\mathrm{sf}_\mathrm{N}}
=\alpha_\mathrm{N}\frac{J_0}{2}
\end{equation}
where Eq.~(\ref{ep2}) has been used. Comparing Eqs.~(\ref{qs}) and (\ref{jleak}), one can identify $r^\mathrm{sf}_\mathrm{F}$ ($r^\mathrm{sf}_\mathrm{N}$) as the resistor in parallel with the SA capacitor $C_\mathrm{sa}^\mathrm{F}$ ($C_\mathrm{sa}^\mathrm{N}$) as shown in Fig.~\ref{fig_circuit_fin}(b). Using $\alpha_\mathrm{F}+\alpha_\mathrm{N}=\beta$, we have
\begin{equation}\label{leakcurrent}
\beta\frac{J_0}{2}=J_\mathrm{sf}^\mathrm{F}+J_\mathrm{sf}^\mathrm{N},
\end{equation}
which can also be derived directly by integrating Eq.~(\ref{jspinderivative2}) from $-\infty$ to $\infty$. In fact, $\beta{J}_0/2$ is half of $J_\mathrm{m}(-\infty)$, which is the spin current density produced by the bulk FM layer. It decreases to zero exponentially at $z=\infty$ after the spin relaxation around the interface. During this process, half of $J_\mathrm{m}(-\infty)$, that is $\beta{J}_0/2$, leaks from spin-down channel to spin-up channel in the FM and NM layers.

%These equations can be interpreted in the following way: the leakage current ($J_\mathrm{sf}$) is the ratio of the effective potential to the characteristic resistance ($r^\mathrm{sf}$).

\subsection{Heat generation}

The spin relaxation also generates heat as the spin-flip scattering causes dissipation of the energy stored in the chemical-potential splitting. The heat generation rate due to spin-flip scattering can be written as
\begin{align}
\Sigma_\mathrm{heat}^\mathrm{F,sf}&=\int_{-\infty}^0\frac{N_{s}\left[\mu_\mathrm{m}(z)\right]^2}
{\tau_\mathrm{sf}^\mathrm{F}}\mathrm{d}z,\\
\Sigma_\mathrm{heat}^\mathrm{N,sf}&=\int_0^\infty\frac{N_{s}\left[\mu_\mathrm{m}(z)\right]^2}
{\tau_\mathrm{sf}^\mathrm{N}}\mathrm{d}z,
\end{align}
for the FM and NM layers, respectively. Using Eqs.~(\ref{deltamuf}) and (\ref{deltamun}), we can write $\Sigma_\mathrm{heat}^\mathrm{F,sf}$ and $\Sigma_\mathrm{heat}^\mathrm{N,sf}$ formally as
\begin{align}
\Sigma_\mathrm{heat}^\mathrm{F,sf}&=r^\mathrm{sf}_\mathrm{F}\left(J_\mathrm{sf}^\mathrm{F}\right)^2
=\frac{1}{2}r_\mathrm{F}\alpha_\mathrm{F}^2J_0^2,\label{leakpf}\\
\Sigma_\mathrm{heat}^\mathrm{N,sf}&=r^\mathrm{sf}_\mathrm{N}\left(J_\mathrm{sf}^\mathrm{N}\right)^2
=\frac{1}{2}r_\mathrm{N}\alpha_\mathrm{N}^2J_0^2,\label{leakpn}
\end{align}
where Eq.~(\ref{jleak}) has been used. Thus the heat generation due to spin relaxation can be modeled as the Joule heat of the resistors in parallel with the SA capacitors.

\subsection{An effective circuit for spin transport}

We have constructed an effective circuit, shown in Fig.~\ref{fig_circuit_fin}(b), to model the spin transport, which takes place on the scales of spin-diffusion length around the interface. We will explain briefly how we build the circuit on the basis of the physical analysis above. Meanwhile, we will also point out the differences between this circuit and those in the Valet-Fert theory.~\cite{vf93}

In Fig.~\ref{fig_circuit_fin}(b), the SA capacitors ($C_\mathrm{sa}^\mathrm{F}$ and $C_\mathrm{sa}^\mathrm{N}$) are connected in parallel with the resistors ($r_\mathrm{F}^\mathrm{sf}$ and $r_\mathrm{N}^\mathrm{sf}$), respectively. The two capacitors model the charge storage in the two spin channels due to spin accumulation as shown by Eq.~(\ref{qs}). The leakage current of the SA capacitors models the electron flow from one spin channel to the other due to spin relaxation as shown by Eqs.~(\ref{jsff}) and (\ref{jsfn}). Comparing Eqs.~(\ref{qs}) and (\ref{jleak}), one can see that the two capacitors have the same voltages as the two resistors, respectively. Therefore, the leakage can be represented equivalently by connecting the two finite resistors in parallel with the SA capacitors, respectively. The voltages here are defined by Eq.~(\ref{ep}) and stand for the effective voltages of the chemical-potential splitting, which may be measured by some spin-resolved optical method. This circuit has one obvious difference from those in the Valet-Fert theory: the capacitors and resistors here are connected between the effective electrodes for the two spin channels instead of the real electrodes.

The current sources of this circuit supply only the spin-polarized part of the current in each spin channel. This feature also makes the circuit different from those in the Valet-Fert theory.~\cite{vf93} There are two sources of the spin-polarized current in the junction: the FM layer with bulk spin asymmetry coefficient $\beta\neq{0}$ and the contact with interfacial spin asymmetry coefficient $\gamma\neq{0}$. They are represented by two independent current sources in the circuit: $\beta{J}_0/2$ and $\gamma{J}_0/2$. The current source $\beta{J}_0/2$ is based on Eq.~(\ref{leakcurrent}) and the explanation following it. The introduction of the current source $\gamma{J}_0/2$ can be understood as follows. If we consider the current driven by the electric field (ef) alone (without spin accumulation) at the spin-selective contact, we have
\begin{equation}
J_\pm^\mathrm{ef}(z_\mathrm{C})=\left(1\mp\gamma\right){J_0}/2,\label{efc}
\end{equation}
for the two spin channels, where $J_0$ flows through $r_+=2r_\mathrm{b}^\ast(1+\gamma)$ and $r_-=2r_\mathrm{b}^\ast(1-\gamma)$ in parallel. Here the spin-up electrons are assumed to be in the minority channel without loss of generality. The spin-selective contact makes $J_\pm^\mathrm{ef}(z_\mathrm{C})$ deviate from its unpolarized value $J_0/2$ by an amount of $\gamma{J}_0/2$, and thus it can be modeled as a spin-current source.

The capacitors are charged by the two current sources. This models the formation of spin accumulation under the drive of the spin current. If the interface resistance is negligible, the two capacitors will be charged simultaneously by the current source $\beta{J}_0/2$ and thus they should be connected in parallel as shown in Fig.~\ref{fig_circuit_fin}(b). When the current source due to the interface is taken into account, we usually hope that it can enhance the voltage (or equivalently spin accumulation) of $C_\mathrm{sa}^\mathrm{N}$. Thus it is reasonable to connect it in the way shown in Fig.~\ref{fig_circuit_fin}(b).

Two additional resistances, $r_\mathrm{F}^\mathrm{sd}$ and $r_\mathrm{N}^\mathrm{sd}$, need to be introduced in the effective circuit to model the heat generation due to the spin diffusion in the FM and NM layers, denoted by $\Sigma_\mathrm{heat}^\mathrm{F,sd}$ and $\Sigma_\mathrm{heat}^\mathrm{N,sd}$, respectively. Using a macroscopic approach based on the Boltzmann equation, we can prove in general that the spin diffusion leads to the same heat generation as the spin relaxation (or the spin-flip scattering) in semi-infinite layers.~\cite{Zhang17cpl} To meet this requirement, we connect $r_\mathrm{F}^\mathrm{sd}$ and $r_\mathrm{N}^\mathrm{sd}$ in series with $r_\mathrm{F}^\mathrm{sf}$ and $r_\mathrm{N}^\mathrm{sf}$, respectively, as shown in Fig.~\ref{fig_circuit_fin}(b). Thus $r_\mathrm{F}^\mathrm{sd}$ and $r_\mathrm{N}^\mathrm{sd}$ should be defined as
\begin{equation}\label{rsd}
r_\mathrm{F}^\mathrm{sd}=r_\mathrm{F}^\mathrm{sf},\qquad r_\mathrm{N}^\mathrm{sd}=r_\mathrm{N}^\mathrm{sf},
\end{equation}
to satisfy $\Sigma_\mathrm{heat}^\mathrm{F,sd}=\Sigma_\mathrm{heat}^\mathrm{F,sf}$ and $\Sigma_\mathrm{heat}^\mathrm{N,sd}=\Sigma_\mathrm{heat}^\mathrm{N,sf}$. Then using Eqs.~(\ref{leakpf}) and (\ref{leakpn}), we can write the spin-dependent heat generation as
\begin{align}
\Sigma_\mathrm{heat}^\mathrm{F}&=\Sigma_\mathrm{heat}^\mathrm{F,sd}+\Sigma_\mathrm{heat}^\mathrm{F,sf}
=r_\mathrm{F}\alpha_\mathrm{F}^2J_0^2,\label{hgf}\\
\Sigma_\mathrm{heat}^\mathrm{N}&=\Sigma_\mathrm{heat}^\mathrm{N,sd}+\Sigma_\mathrm{heat}^\mathrm{N,sf}
=r_\mathrm{N}\alpha_\mathrm{N}^2J_0^2,\label{hgn}
\end{align}
for the FM and NM layers, respectively.

%The heat generation due to the spin diffusion can be written as
%\begin{align}
%\Sigma_\mathrm{heat}^\mathrm{F,sd}&=r^\mathrm{sd}_\mathrm{F}\left(J_\mathrm{sf}^\mathrm{F}\right)^2
%=\frac{1}{2}r_\mathrm{F}\alpha_\mathrm{F}^2J_0^2\label{diffusionf}\\
%\Sigma_\mathrm{heat}^\mathrm{N,sd}&=r^\mathrm{sd}_\mathrm{N}\left(J_\mathrm{sf}^\mathrm{N}\right)^2
%=\frac{1}{2}r_\mathrm{N}\alpha_\mathrm{N}^2J_0^2\label{diffusionn}
%\end{align}
%which are equal to $\Sigma_\mathrm{heat}^\mathrm{F,sf}$ and $\Sigma_\mathrm{heat}^\mathrm{N,sf}$, respectively.

The currents $J_\mathrm{sf}^\mathrm{F}$ and $J_\mathrm{sf}^\mathrm{N}$ given by Eq.~(\ref{jleak}) can also be derived by applying the superposition theorem of electrical circuits to the circuit in Fig.~\ref{fig_circuit_fin}(b). Both current sources drive current to flow from the spin-down to the spin-up channel in the NM layer. However, they drive opposite currents in the FM layer and the net current $J_\mathrm{sf}^\mathrm{F}$ can be negative in some cases. The effective voltage on the resistance $r_\mathrm{F}^\mathrm{sd}$ ($r_\mathrm{N}^\mathrm{sd}$) is also equal to that on $r_\mathrm{F}^\mathrm{sf}$ ($r_\mathrm{N}^\mathrm{sf}$), which is consistent with Eq.~(\ref{ep}). Moreover, using Kirchhoff's current law, one can calculate the current density flowing through the resistance $4r_\mathrm{b}^\ast$
\begin{equation}
J_\mathrm{sd}^\mathrm{C}=\gamma\frac{J_0}{2}-J_\mathrm{sf}^\mathrm{N}
=\alpha_\mathrm{C}\frac{J_0}{2},\label{jc}
\end{equation}
where we have introduced the dimensionless parameter
\begin{equation}\label{alphac}
\alpha_\mathrm{C}=\gamma-\alpha_\mathrm{N}
=\frac{\gamma(r_\mathrm{F}+r_\mathrm{N})-\beta{r}_\mathrm{F}}
{r_\mathrm{F}+r_\mathrm{b}^\ast+r_\mathrm{N}}.
\end{equation}
Then the effective voltage on the resistance $4r_\mathrm{b}^\ast$ can be written as $2(V_\mathrm{c}^\mathrm{N}-V_\mathrm{c}^\mathrm{F})$ or $[\mu_\mathrm{m}(0^+)-\mu_\mathrm{m}(0^-)]/e$. Using Joule's law, we can write the heat generation of the resistance $4r_\mathrm{b}^\ast$ as
\begin{equation}\label{leakpb}
\Sigma_\mathrm{heat}^\mathrm{C}=4r_\mathrm{b}^\ast\left(J_\mathrm{sd}^\mathrm{C}\right)^2
=\frac{\left(2V_\mathrm{c}^\mathrm{N}-2V_\mathrm{c}^\mathrm{F}\right)^2}{4r_\mathrm{b}^\ast}
=r_\mathrm{b}^\ast\alpha_\mathrm{C}^2J_0^2,
\end{equation}
which is equal to the result derived by using a more microscopic approach.~\cite{Zhang17pb} It has been shown, in Ref.~\onlinecite{Zhang17pb}, that $\Sigma_\mathrm{heat}^\mathrm{C}$ results only from the spin diffusion across the interface because the spin relaxation at the interface is negligible.~\cite{vf93} Combining Eqs.~(\ref{hgf}), (\ref{hgn}), and (\ref{leakpb}), we can write the total heat generation due to the spin relaxation and the spin diffusion as
\begin{equation}\label{hgfn}
\Sigma_\mathrm{heat}^\mathrm{FN}=\Sigma_\mathrm{heat}^\mathrm{F}
+\Sigma_\mathrm{heat}^\mathrm{N}+\Sigma_\mathrm{heat}^\mathrm{C}
=r_\mathrm{FN}^\ast{J}_0^2,
\end{equation}
where $r_\mathrm{FN}^\ast$ is defined as $r_\mathrm{FN}^\ast=r_\mathrm{F}\alpha_\mathrm{F}^2+r_\mathrm{N}\alpha_\mathrm{N}^2 +r_\mathrm{b}^\ast\alpha_\mathrm{C}^2$. One can verify that $r_\mathrm{FN}^\ast$ is equal to the spin-coupled interface resistance of the FM/NM junction.~\cite{vf93,Zhang17pb}

\begin{figure}
\includegraphics[width=0.48\textwidth]{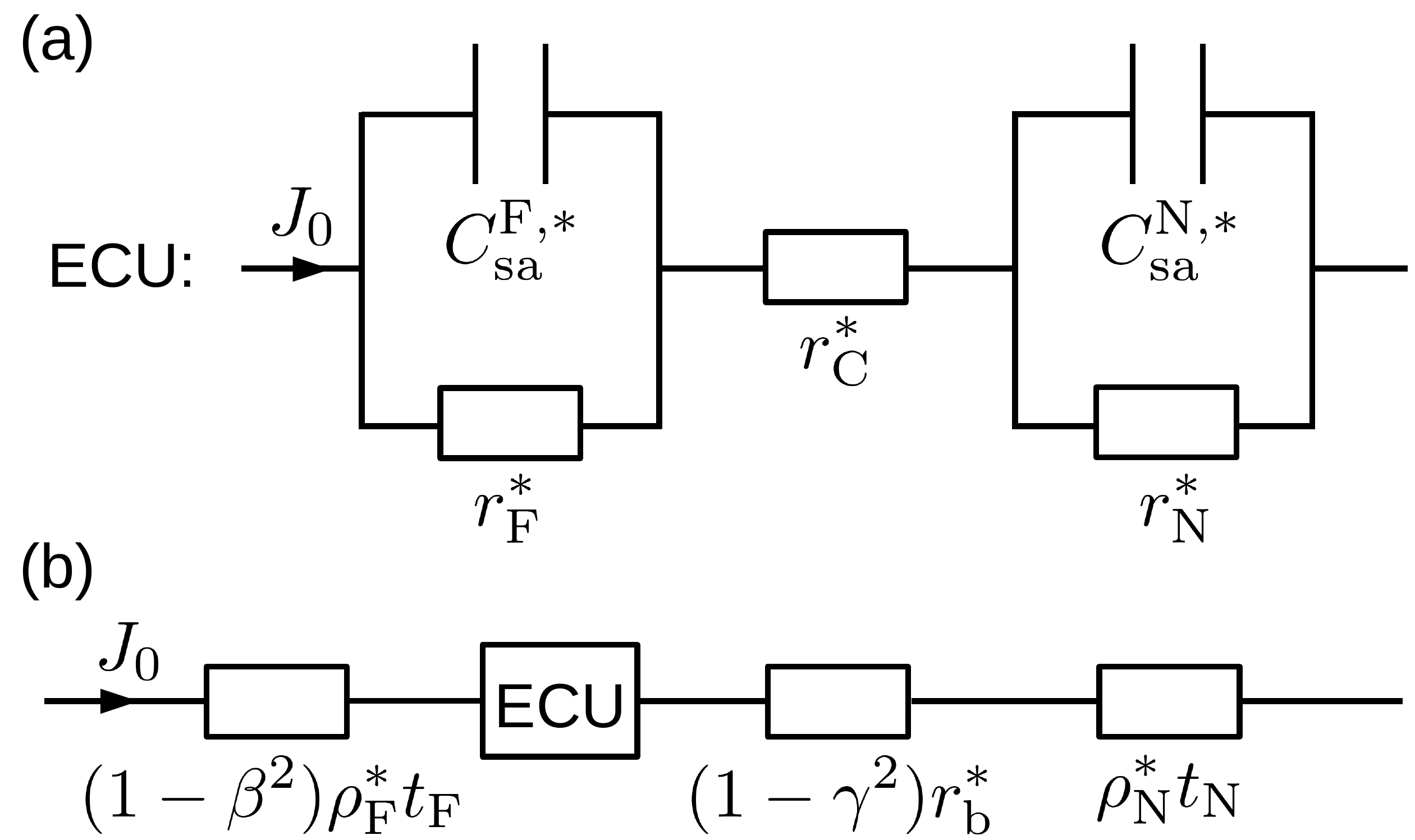}
\caption{\label{fig_circuit_fin2} (a) An equivalent circuit unit (ECU) for the ECST shown in Fig.~\ref{fig_circuit_fin}(b). In the ECU, the current flowing through the various resistances is the total charge current $J_0$, instead of the spin currents in ECST. The equivalent resistances $r_\mathrm{F}^\ast$, $r_\mathrm{N}^\ast$, and $r_\mathrm{C}^\ast$ are given in Eqs.~(\ref{req}) and (\ref{reqc}), respectively. The ESA capacitances $C_\mathrm{sa}^{\mathrm{F},\ast}$ and $C_\mathrm{sa}^{\mathrm{N},\ast}$ are defined in Eq.~(\ref{ceq}). (b) An overall equivalent circuit of the FM/NM junction shown in Fig.~\ref{fig_circuit_fin}(a). The block labeled by ``ECU'' stands for the circuit unit in Fig.~\ref{fig_circuit_fin2}(a). The FM (NM) layer thickness $t_\mathrm{F}$ ($t_\mathrm{N}$) is much larger than the spin diffusion length $l_\mathrm{sf}^\mathrm{F}$ ($l_\mathrm{sf}^\mathrm{N}$).}
\end{figure}

\subsection{An overall equivalent circuit}

To make our results comparable with Rashba's theory and experimental results, we have also constructed an overall equivalent circuit, which includes the ECST shown in Fig.~\ref{fig_circuit_fin}(b). In the equivalent circuit shown in Fig.~\ref{fig_circuit_fin2}, the current source is the total charge current $J_0$ instead of the spin current sources $\beta{J}_0/2$ and $\gamma{J}_0/2$. This makes the circuit similar to those in the Valet-Fert theory. However, we have combine the two spin channels to make the circuit more comparable to the practical situation. The two terminals in Fig.~\ref{fig_circuit_fin2}(b) are supposed to be connected to the real electrodes, which are not shown for simplicity. Thus its voltage can be measured directly in experiments.

We require that the equivalent circuit has the same energy storage $W_\mathrm{sa}^\mathrm{F(N)}$ and heat generation $\Sigma_\mathrm{sa}^\mathrm{F(N)}$ as the ECST, which in turn has the same $W_\mathrm{sa}^\mathrm{F(N)}$ and $\Sigma_\mathrm{sa}^\mathrm{F(N)}$ as the real physical system. We first determine the equivalent resistances by requiring the same heat generation. Using Eqs.~(\ref{hgf}) and (\ref{hgn}), we can set the equivalent resistances to be
\begin{equation}\label{req}
r_\mathrm{F}^\ast=r_\mathrm{F}\alpha_\mathrm{F}^2,\qquad
r_\mathrm{N}^\ast=r_\mathrm{N}\alpha_\mathrm{N}^2,
\end{equation}
for the FM and NM layers, respectively. Moreover, using Eq.~(\ref{leakpb}), we can set the equivalent resistance for $4r_\mathrm{b}^\ast$ in Fig.~\ref{fig_circuit_fin}(b) to be
\begin{equation}\label{reqc}
r_\mathrm{C}^\ast=r_\mathrm{b}^\ast\alpha_\mathrm{C}^2,
\end{equation}
which satisfies $r_\mathrm{FN}^\ast=r_\mathrm{F}^\ast+r_\mathrm{N}^\ast+r_\mathrm{C}^\ast$. Then the equivalent voltages on them can be written as
\begin{equation}\label{veq}
V_\mathrm{F}^\ast=r_\mathrm{F}^\ast{J}_0,\qquad
V_\mathrm{N}^\ast=r_\mathrm{N}^\ast{J}_0.
\end{equation}
For simplicity, we assume that the equivalent spin-accumulation (ESA) capacitance $C_\mathrm{sa}^{\mathrm{F},\ast}$ ($C_\mathrm{sa}^{\mathrm{N},\ast}$) is connected in parallel with the equivalent resistance $r_\mathrm{F}^\ast$ ($r_\mathrm{N}^\ast$). Then, $V_\mathrm{F}^\ast$ ($V_\mathrm{N}^\ast$) is also the voltage on $C_\mathrm{sa}^{\mathrm{F},\ast}$ ($C_\mathrm{sa}^{\mathrm{N},\ast}$). We require that $C_\mathrm{sa}^{\mathrm{F},\ast}$ and $C_\mathrm{sa}^{\mathrm{N},\ast}$ have the same energy storage as $C_\mathrm{sa}^\mathrm{F}$ and $C_\mathrm{sa}^\mathrm{N}$
\begin{equation}\label{esceq}
W_\mathrm{sa}^\mathrm{F}=\frac{1}{2}C_\mathrm{sa}^{\mathrm{F},\ast}
\left(V_\mathrm{F}^\ast\right)^2,\qquad
W_\mathrm{sa}^\mathrm{N}=\frac{1}{2}C_\mathrm{sa}^{\mathrm{N},\ast}
\left(V_\mathrm{N}^\ast\right)^2.
\end{equation}
Substituting Eqs.~(\ref{esc}) and (\ref{veq}) into Eq.~(\ref{esceq}), we can find
\begin{equation}\label{ceq}
C_\mathrm{sa}^{\mathrm{F},\ast}=\frac{T_1^\mathrm{F}}{2r_\mathrm{F}^\ast}
=\frac{C_\mathrm{sa}^\mathrm{F}}{\alpha_\mathrm{F}^2},\qquad
C_\mathrm{sa}^{\mathrm{N},\ast}=\frac{T_1^\mathrm{N}}{2r_\mathrm{N}^\ast}
=\frac{C_\mathrm{sa}^\mathrm{N}}{\alpha_\mathrm{N}^2},
\end{equation}
for the FM and NM layers, respectively. Finally, we can construct an overall equivalent circuit shown in Fig.~\ref{fig_circuit_fin2}(b), which also include the normal resistances $(1-\beta^2)\rho_\mathrm{F}^\ast{t}_\mathrm{F}$, $(1-\gamma^2)r_\mathrm{b}^\ast$, and $\rho_\mathrm{N}^\ast{t}_\mathrm{N}$.

\section{Reinterpretation of the magnetoimpedance\label{discussion}}
%objective and motivation; quasi-static approximation
% The imaginary part of the magneto-impedance, the reactance $X_\mathrm{FN}$, will be rewritten in terms of the SA or ESA capacitances, and their connections to the diffusion capacitance will also be discussed.

As an application of the SA or ESA capacitances, we revisit the magnetoimpedance of the FM/NM junction by using the equivalent circuit shown in Fig.~\ref{fig_circuit_fin2}. The reactance in this configuration has been interpreted by Rashba in terms of the diffusion capacitance $C_\mathrm{diff}$.~\cite{Rashba02a}

The SA capacitance is approximately a constant for time-varying DC and (sinusoidal) AC signals in the regime for the quasi-static approximation to be valid. When the junction is driven by a constant DC current as discussed in the previous section, the spin accumulation is an exponential function in each layer and the SA capacitance is a constant independent on the current density. In general, the SA capacitance will change if the driving current becomes time-dependent, for example, a time-varying DC or AC signal. However, if the charge current $J(t)$ changes with time slowly enough that the solutions to the time-dependent equations are still exponential functions approximately at each moment, they can be replaced by the static solutions for the DC current of the same density $J(t)$. Then the SA capacitance can be assumed to be a constant for simplicity. This is the quasi-static approximation, which has been widely used in circuit analysis and also applied to the diffusion capacitance of a p-n junction.~\cite{Neamen2012}

The valid range for the quasi-static approximation can be determined as follows. Since we are considering the impedance, it is enough to focus on the sinusoidal AC signal. The characteristic time scale of the spin dynamics is the spin relaxation time $T_1^\mathrm{N}$, if we have the relation $T_1^\mathrm{N}>T_1^\mathrm{F}$ as usual.~\cite{zhu08,zhang02} Therefore, the quasi-static approximation is valid if the frequency of the driving current satisfies $\omega\ll{1/T_1^\mathrm{N}}$ or equivalently $\omega{C}_\mathrm{sa}^{\mathrm{N},\ast}\ll{1/\left(2r_\mathrm{N}^\ast\right)}$. In practice, most experiments on time-dependent transport use slowly-varying or low-frequency AC signals.~\cite{Xiao01,Chien2006} In the high-frequency regime, the quasi-static approximation breaks down because the spin dynamics cannot follow the variation of the driving signal and the spin accumulation deviates from the exponential form severely.~\cite{zhu08} In this case, the spin wave-diffusion theory should be used instead. Consequently, the inductive terms becomes remarkable and may be dominant over the capacitive terms if the frequency is high enough.~\cite{zhu09}

%In the low-frequency regime, Eqs.~(10) and (11) in Ref.~\onlinecite{zhu08} reduce to the conventional spin-diffusion equations, where $-\tau\partial{J}_{\rm{m}}(z,t)/\partial{t}$ and the space-dependence of $J(z,t)$ have been neglected. We also need the equations for the charge dynamics, where $-\tau\partial{J(z,t)}/\partial{t}$ has been neglected.

%Thus we can figure out first the capacitance of the structure driven by a DC and then calculate the impedance with an AC drive by constructing equivalent circuit.

% in order to find the origin of its imaginary part, that is, the reactance

If the AC current density is set to be $J(t)=J_0\exp\left(-\mathrm{i}\omega{t}\right)$, the impedances of $C_\mathrm{sa}^{\mathrm{F},\ast}$ and $C_\mathrm{sa}^{\mathrm{N},\ast}$ in Fig.~\ref{fig_circuit_fin2}(a) can be written, respectively, as $\mathrm{i}/\left(\omega{C}_\mathrm{sa}^{\mathrm{F},\ast}\right)$ and $\mathrm{i}/\left(\omega{C}_\mathrm{sa}^{\mathrm{N},\ast}\right)$ using the quasi-static approximation. Using usual circuit theorems, we can calculate the impedance, $Z_\mathrm{ECU}$, of the ECU in Fig.~\ref{fig_circuit_fin2}(a). Then expanding $Z_\mathrm{ECU}$ in terms of $\omega$ and keeping up to the first-order terms, we have
\begin{equation}\label{zecu}
Z_\mathrm{ECU}=r_\mathrm{FN}^\ast+\mathrm{i}X_\mathrm{sa}^\mathrm{FN},
\end{equation}
where $r_\mathrm{FN}^\ast$ is defined in Eq.~(\ref{hgfn}) and the reactance $X_\mathrm{sa}^\mathrm{FN}$ can be written as
\begin{equation}\label{xfn}
X_\mathrm{sa}^\mathrm{FN}=\omega{C}_\mathrm{sa}^{\mathrm{F},\ast}\left(r_\mathrm{F}^\ast\right)^2
+\omega{C}_\mathrm{sa}^{\mathrm{N},\ast}\left(r_\mathrm{N}^\ast\right)^2.
\end{equation}

%\begin{equation}
%r_\mathrm{SI}^\mathrm{FN}(\omega)=\frac{r_\mathrm{F}^\ast}
%{1+\left(r_\mathrm{F}^\ast{C}_\mathrm{sa}^{\mathrm{F},\ast}\omega\right)^2}
%+\frac{r_\mathrm{N}^\ast}
%{1+\left(r_\mathrm{N}^\ast{C}_\mathrm{sa}^{\mathrm{N},\ast}\omega\right)^2}
%+r_\mathrm{C}^\ast
%\end{equation}

We will prove that the impedance given in Eq.~(\ref{zecu}) is equal to that derived by Rashba ($\mathcal{Z}_\mathrm{n-eq}$) in the low-frequency regime.~\cite{Rashba02a} Rashba's basic results are outlined with our notations in Appendix~\ref{fnmi}. Using Eqs.~(\ref{alphafn}) and (\ref{alphac}) in the present paper, we can rewrite the real part of the impedance, $R_\mathrm{n-eq}$ given in Eq.~(\ref{rneq}), in a form with more transparent physical interpretation
\begin{equation}
R_\mathrm{n-eq}=r_\mathrm{F}^\ast+r_\mathrm{N}^\ast+r_\mathrm{C}^\ast=r_\mathrm{FN}^\ast,
\end{equation}
which is equal to the real part of $Z_\mathrm{ECU}$ in Eq.~(\ref{zecu}). The imaginary part of the impedance, $\mathrm{Im}(\mathcal{Z}_\mathrm{n-eq})$ in Eq.~(\ref{xfnrashba}), can be rewritten as
\begin{equation}\label{xfn2} \mathrm{Im}(\mathcal{Z}_\mathrm{n-eq})=\omega\frac{T_1^\mathrm{F}}{2r_\mathrm{F}}
(r_\mathrm{F}\alpha_\mathrm{F})^2+
\omega\frac{T_1^\mathrm{N}}{2r_\mathrm{N}}(r_\mathrm{N}\alpha_\mathrm{N})^2,
\end{equation}
where we have used Eqs.~(\ref{alphafn}) and (\ref{alphac}). Then by using the SA capacitances defined in Eq.~(\ref{sc}), we can rewrite Eq.~(\ref{xfn2}) as
\begin{equation}\label{xfns}
\mathrm{Im}(\mathcal{Z}_\mathrm{n-eq})=\omega{C}_\mathrm{sa}^\mathrm{F}
\left(r_\mathrm{F}\alpha_\mathrm{F}\right)^2
+\omega{C}_\mathrm{sa}^\mathrm{N}\left(r_\mathrm{N}\alpha_\mathrm{N}\right)^2.
\end{equation}
Finally, by using Eqs.~(\ref{req}) and (\ref{ceq}), we can prove that $\mathrm{Im}(\mathcal{Z}_\mathrm{n-eq})$ is equal to $X_\mathrm{sa}^\mathrm{FN}$ in Eq.~(\ref{xfn}). Therefore, the impedance $Z_\mathrm{ECU}$ is the same as $\mathcal{Z}_\mathrm{n-eq}$ derived by Rashba and the ECU shown in Fig.~\ref{fig_circuit_fin2}(a) can also be used to interpret $\mathcal{Z}_\mathrm{n-eq}$.

%In the low-frequency limit, $\omega\to0$, the reactance of the circuit unit composed of a capacitance $C$ and a resistance $R$ in parallel can be written as $X=\omega{C}R^2$. The energy stored in the capacitor is $CU^2/2=XJ^2/(2\omega)$ in the constant DC situation. Therefore, under DC drive $J_0$, the energy storage associated with the reactance of the FM/NM junction can be written as
%\begin{equation}
%\frac{X^\mathrm{FN}J_0^2}{2\omega}=
%\frac{1}{2}\left[C_\mathrm{eq}^\mathrm{F}\left(V_\mathrm{eq}^\mathrm{F}\right)^2+
%C_\mathrm{eq}^\mathrm{N}\left(V_\mathrm{eq}^\mathrm{N}\right)^2\right]
%\end{equation}
%where we have used Eqs.~(\ref{veq}) and (\ref{ceq}). This energy is
%equal to the sum of $W_\mathrm{eq}^\mathrm{F}$ and
%$W_\mathrm{eq}^\mathrm{N}$, which is also the same as $W_\mathrm{sc}$,
%the energy stored in the spin capacitors. This makes us believe that
%the energy storage associated with reactance $X^\mathrm{(FN)}$ is the
%splitting energy due to spin accumulation, which is essentially
%kinetic energy instead of electrostatic energy.

In comparison to Rashba's theory, we wrote the reactance of the FM/NM junction in terms of the SA or ESA capacitances instead of the diffusion capacitance $C_\mathrm{diff}$ defined in Eq.~(\ref{cdiff}). One benefit is that the form of the SA or ESA capacitances resemble obviously that of the diffusion capacitance in a p-n junction. More importantly, our results revealed the form of the energy storage associated with the reactance. In the low-frequency limit, the energy storage associated with $\mathcal{Z}_\mathrm{n-eq}$ is given by Eq.~(\ref{esc}) or (\ref{esceq}), which is the splitting energy of the spin-dependent chemical potentials instead of the electrostatic energy. Our results also suggest a way to calculate the reactance without using the time-dependent equations.

%In practice, what really matters is the reactance instead of the capacitance.

One characteristic of the reactance, $X_\mathrm{sa}^\mathrm{FN}$, is that it is zero if $\beta=0$ and $\gamma=0$. This means that a spin-current source is indispensable to the reactance we consider. Multilayers with $\beta=0$ and $\gamma=0$ may still have parasitic capacitance in parallel with resistance as long as the material resistivity varies from layer to layer. It results from the induced charge accumulation at the interfaces, which adjusts the electric field in different layers to maintain a homogeneous current across the entire multilayer. The energy stored in this kind of capacitance is essentially electrostatic energy and it will not be discussed in this paper.

It is worthwhile to compare the magnetoimpedance in the present paper with that reviewed in Ref.~\onlinecite{Phan2008}. Although the same term ``magnetoimpedance'' is used in the two papers, their configurations are different in the following three ways. First, Ref.~\onlinecite{Phan2008} discussed the change of impedance with an applied magnetic field, whereas such an external field is not essential in our problem and the change of impedance here depends on the spin polarization as shown by Eq.~(\ref{xfns}). Second, the impedance in Ref.~\onlinecite{Phan2008} is mostly inductive. On the contrary, our impedance is capacitive. Finally, Ref.~\onlinecite{Phan2008} was mainly concerned with the impedance of a single FM conductor. However, we are interested in magnetic multilayers, such as an FM/NM junction. Therefore, the imaginary part of the magnetoimpedance reviewed in Ref.~\onlinecite{Phan2008} arises from quite different origin and is beyond the scope of the current paper.

% Integrating from $-\infty$ to $\infty$ on both sides of
% Eq.~(\ref{new3}), we have
% \begin{equation}
%   \beta\frac{J(t)}{2}=\frac{\partial{Q}_\mathrm{sc}(t)}{\partial{t}}
%   +J_\mathrm{r}(t)
% \end{equation}

\section{Conclusions\label{conclusion}}

In summary, we model the two spin channels of a metal or degenerate semiconductor as the two plates of a capacitor. By using a method similar to that for studying quantum capacitance, we show that the SA capacitance can be introduced for each layer to measure its ability to store spins rather than charge and a spatial charge storage is not essential to it. The energy stored in the SA capacitors is the splitting energy of the spin-dependent chemical potentials instead of the electrostatic energy. The leakage of the SA capacitor models the spin relaxation. The heat generation due to the spin relaxation can be written formally as the Joule heat of the resistance in parallel with the SA capacitor. Equivalent circuits can be constructed to give a transparent interpretation of the model for a typical FM/NM junction.

As an application of the model, we derive the magnetoimpedance of the FM/NM junction directly from the equivalent circuit using the quasi-static approximation. We rewrite the reactance in terms of the SA capacitances, whose formula is obviously similar to that of the diffusion capacitance in a p-n junction. Our results also reveal the form of the energy storage associated with the reactance. We expect that the SA capacitance can be generalized to other structures, such as MTJs, and used to interpret the ongoing experiments on the magnetoimpedance.

%Our results yield the same value as those found by Rashba using a time-dependent approach and the diffusion capacitance.

\begin{acknowledgments}
We thank Prof.~H. C. Schneider and Y. Suzuki for fruitful discussions. This work was supported by the National Natural Science Foundation of China under Grant Nos~11404013, 11605003, 61405003, and 11474012.
\end{acknowledgments}

%and Beijing Natural Science Foundation (grant number~1112007).
% 1.Yao-Hui Zhu; 2. Jian Liu (11605003); 3. Pei-Song He (61405003); 4. Bao-He Li (11174020); 5. Jun-Jie Shi (11474012)

%This project was supported by National Natural Science Foundation of China (Grant No. 11404013, 11474012, 11174020, 61405003), Beijing Natural Science Foundation (Grant No. 1112007), and Scientific Research Common Program of Beijing Municipal Commission of Education (Grant No. KM201310011008).

\appendix

\section{The FM/NM junction with a constant DC\label{appfn}}

According to the Valet-Fert theory,~\cite{vf93} $\Delta\mu(z)$ and current density $J_\pm(z)$ satisfy the following equations
\begin{align}
&\frac{e}{\sigma_\pm}\frac{\partial{J}_\pm}{\partial{z}} =\pm2\frac{\Delta\mu}{l_\pm^2}\label{spin-flip2},\\
&J_\pm=\sigma_\pm\left(F\pm\frac{1}{e}
\frac{\partial\Delta\mu}{\partial{z}}\right),\label{Ohm's law2}
\end{align}
where $\sigma_\pm$ and $l_\pm$ denote the conductivity and spin-diffusion length for spin ``$\pm$'' (``up'' or ``down''), respectively. The field $F$ is defined as $F=(1/e)\partial\bar\mu/\partial{z}$, where $\bar\mu=(\bar\mu_{+}+\bar\mu_{-})/2$ is the average electrochemical potential.

% In FM layers, the conductivity $\sigma_\pm$ satisfies the relation $1/\sigma_{\uparrow(\downarrow)}=2\rho_\mathrm{F}^\ast[1-(+)\beta]$, where the subscript ``$\uparrow$'' (``$\downarrow$'') denotes the majority (minority) spin direction.

%In Eq.~(\ref{new5a}), the bulk asymmetry parameter $\tilde\beta$ is $\beta$ in FM layer and zero in NM layer. In FM layers with `up' magnetization, $\beta$ is defined by following relation \begin{equation}\label{rhopm}
%\rho_\pm=2\rho_\mathrm{F}^\ast(1\pm\beta)
%\end{equation}
%where $\rho_\mathrm{F}^\ast$ is the effective resistivity of the FM
%layer.

Several useful equations can be derived from Eqs.~(\ref{spin-flip2}) and (\ref{Ohm's law2}). Subtracting the ``$\pm$'' components of Eq.~(\ref{spin-flip2}) directly, we have
\begin{equation}\label{jspinderivative1}
\frac{\partial{J}_\mathrm{m}}{\partial{z}}
=\frac{\Delta\mu}{er_\mathrm{F(N)}l_\mathrm{sf}^\mathrm{F(N)}},
\end{equation}
where we have introduced the spin current density $J_\mathrm{m}(z)=J_{+}(z)-J_{-}(z)$. The spin diffusion length is defined by $1/l_\mathrm{sf}^2=1/l_{+}^2+1/l_{-}^2$. The resistance $r_\mathrm{F}$ ($r_\mathrm{N}$) is defined in Eq.~(\ref{rfn}). Multiplying $\sigma_\pm$ on both sides of Eq.~(\ref{spin-flip2}) and subtracting its ``$\pm$'' components, we have
\begin{equation}\label{jspinderivative2}
\frac{\partial{J}_\mathrm{m}}{\partial{z}}
=\frac{4eN_s\Delta\mu}{\tau_\mathrm{sf}^\mathrm{F(N)}},
\end{equation}
where $\sigma_\pm=e^2N_sD_\pm$ and $l_\pm^2=D_\pm\tau_\mathrm{sf}$ have been used.~\cite{zhu08} Moreover, $D_\pm$ is the diffusion constant for spin ``$\pm$''. Comparing Eqs.~(\ref{jspinderivative1}) and (\ref{jspinderivative2}), we can find the relation
\begin{equation}\label{csa}
\frac{T_1^\mathrm{F(N)}}{2r_\mathrm{F(N)}}=e^2N_sl_\mathrm{sf}^\mathrm{F(N)},
\end{equation}
where $\tau_\mathrm{sf}^\mathrm{F(N)}=2T_1^\mathrm{F(N)}$ has been used. Dividing both sides of Eq.~(\ref{Ohm's law2}) by $\sigma_\pm$ and then subtracting its `$\pm$' components, we have
\begin{equation}\label{spincurrent}
J_\mathrm{m}=\mp\beta{J}+\frac{1}{e\rho_\mathrm{F}^\ast}
\frac{\partial\Delta\mu}{\partial{z}}
\end{equation}
for FM layers. Here, ``$-$'' (``$+$'') corresponds to ``up'' (``down'') magnetization of the FM layer. For NM layers, one only needs to set $\beta=0$ and replace $\rho_\mathrm{F}^\ast$ by $\rho_\mathrm{N}^\ast$ in Eq.~(\ref{spincurrent}). Substituting Eq.~(\ref{spincurrent}) into Eq.~(\ref{jspinderivative1}), we can derive Eq.~(\ref{new6a}) in section~\ref{theory}.

At the interface located at $z=0$, the electrochemical potential $\bar\mu_\pm(z)$ and $J_\pm(z)$ satisfy the boundary conditions
\begin{align}
&J_{s}\left(0^{+}\right)-J_{s}\left(0^{-}\right)=0,\label{bcif1}\\
&\bar{\mu}_{s}\left(0^{+}\right)
-\bar{\mu}_{s}\left(0^{-}\right)
=er_{s}J_{s}\left(0\right)\label{bcif2},
\end{align}
where $r_{s}$ and $\gamma$ are defined in Eq.~(\ref{bcr}). Then applying the Valet-Fert theory to the FM/NM junction considered in section~\ref{theory}, we have
\begin{align}
&\Delta\mu(z)=\Delta\mu(0^{-})\exp(z/l_\mathrm{sf}^\mathrm{F}),\qquad\>\>\;{z<0}\label{deltamuf}\\
&\Delta\mu(z)=\Delta\mu(0^{+})\exp(-z/l_\mathrm{sf}^\mathrm{N}),\qquad{z>0}\label{deltamun}
\end{align}
where we have introduced
\begin{align}\label{mu0pm}
\Delta\mu(0^{-})=er_\mathrm{F}\alpha_\mathrm{F}J_0,\qquad
\Delta\mu(0^{+})=er_\mathrm{N}\alpha_\mathrm{N}J_0.
\end{align}
%Correspondingly, we also have
%\begin{align}
%&\mu_\mathrm{m}(0^{-})=2\Delta\mu(0^{-})=2er_\mathrm{F}\alpha_\mathrm{F}J_0\\
%&\mu_\mathrm{m}(0^{+})=2\Delta\mu(0^{+})=2er_\mathrm{N}\alpha_\mathrm{N}J_0
%\end{align}
The parameters $\alpha_\mathrm{F}$ and $\alpha_\mathrm{N}$ are defined in Eq.~(\ref{alphafn}).

Substituting Eq.~(\ref{deltamuf}) into Eq.~(\ref{chargefn}), we have
\begin{equation}\label{qsf}
Q_\mathrm{sa}^\mathrm{F}=e^2{N}_{s}l_\mathrm{sf}^\mathrm{F}V_\mathrm{c}^\mathrm{F}
\end{equation}
where $V_\mathrm{c}^\mathrm{F}$ is defined in Eq.~(\ref{ep}). Substituting Eq.~(\ref{csa}) into Eq.~(\ref{qsf}), we find
\begin{equation}
Q_\mathrm{sa}^\mathrm{F}=C_\mathrm{sa}^\mathrm{F}V_\mathrm{c}^\mathrm{F}
\end{equation}
where $C_\mathrm{sa}^\mathrm{F}=T_1^\mathrm{F}/(2r_\mathrm{F})$ is the same as that defined in Eq.~(\ref{sc}). One can also rewrite $Q_\mathrm{sa}^\mathrm{N}$ in a similar way.

\section{Quasi-neutrality approximation\label{qnd}}

The average chemical potential $\mu(z)$ can be simplified by using the quasi-neutrality approximation. The nonequilibrium charge density of each spin channel, $\rho_\pm(z)$, has two exponential terms with different decaying lengths: the Thomas-Fermi screening length and the spin diffusion length.~\cite{zhu14} The two terms will be discussed separately in the following.

The screening length is on the order of angstrom and the charge accumulation of this length scale is not spin-polarized. Therefore, the screening charge can be considered as an interfacial charge layer, which gives rise to a constant electric field on each side of the interface. It leads to a parasitic capacitance that is beyond the scope of the present paper and thus the screening charge becomes irrelevant to our problem.

On the scale of the spin diffusion length, one spin channel has an excess of electrons while the other is deficient in electrons. The excess and deficiency cancel each other exactly in the NM layer, but they have slightly different magnitude in the FM layer, which results in a net charge accumulation coexisting with the spin accumulation. The ratio of the net electron concentration to the excess of electrons in one spin channel (or deficiency in the other) satisfies $\beta\lambda_\mathrm{TF}^2/[(l_\mathrm{sf}^\mathrm{F})^2-\lambda_\mathrm{TF}^2]\ll{1}$ for typical ferromagnetic metals. This net charge accumulation can be neglected for some problems, such as the SA capacitance, in which only the spin-dependent charge densities and the associated chemical potentials really matter. On the other hand, this net charge accumulation also leads to an exponentially-decaying electric field, which in turn results in an extra electric potential. It has been proven that the electric potential energy is much larger than the chemical potential at a certain position and dominates the electrochemical potential.~\cite{zhu14} Therefore, the net charge accumulation is necessary for some problems, such as the interpretation of the GMR effects, where the extra potential plays a crucial role.~\cite{vf93} This interpretation is essentially the widely-used quasi-neutrality approximation.

%In other words, the net charge accumulation is negligible when the problems involve only its nonequilibrium chemical potential.

% This charge accumulation can also lead to a geometrical capacitance, which is beyond the scope of this paper.

\section{Rashba's results for the magnetoimpedance\label{fnmi}}

According to Eqs.~(12) and (13) of Rashba's paper,~\cite{Rashba02a} the nonequilibrium impedance of the FM/NM junction can be rewritten as
\begin{equation}
\begin{split}
\mathcal{Z}_\mathrm{n-eq}(\omega)=&\frac{1}{r_\mathrm{FN}(\omega)}
\big[r_\mathrm{N}(\omega)r_\mathrm{b}^\ast\gamma^2 +r_\mathrm{N}(\omega)r_\mathrm{F}(\omega)\beta^2 \\
&+r_\mathrm{b}^\ast{r}_\mathrm{F}(\omega)\left(\gamma-\beta\right)^2\big]
\end{split}
\end{equation}
where $r_\mathrm{FN}(\omega)$ is defined as $r_\mathrm{FN}(\omega)=r_\mathrm{F}(\omega)+r_\mathrm{N}(\omega)+r_\mathrm{b}^\ast$. To simplify the expression, we have substituted $\gamma$ and $\beta$ for $\Delta\Sigma/\Sigma$ and $\Delta\sigma/\sigma_F$ (used by Rashba~\cite{Rashba02a}), respectively. The resistance $r_\mathrm{b}^\ast$ is the same as $r_c$ in Ref.~\onlinecite{Rashba02a}. The frequency-dependent $r_\mathrm{F(N)}(\omega)$ is defined as
\begin{equation}
r_\mathrm{F}(\omega)=\frac{r_\mathrm{F}}{\sqrt{1-\mathrm{i}\omega{T}_1^\mathrm{F}}},\qquad r_\mathrm{N}(\omega)=\frac{r_\mathrm{N}}{\sqrt{1-\mathrm{i}\omega{T}_1^\mathrm{N}}},
\end{equation}
where $T_1^\mathrm{F}$ ($T_1^\mathrm{N}$) is the same as $\tau_s^F$ ($\tau_s^N$) in Ref.~\onlinecite{Rashba02a}. Expanding $\mathcal{Z}_\mathrm{n-eq}(\omega)$ in $\omega$ and keeping up to the first-order terms, one can write its real part as
\begin{align}\label{rneq}
%\mathrm{Re}\left(\mathcal{Z}_\mathrm{n-eq}\right)
R_\mathrm{n-eq}=\frac{r_\mathrm{N}r_\mathrm{b}^\ast\gamma^2 +r_\mathrm{N}r_\mathrm{F}\beta^2+r_\mathrm{b}^\ast{r}_\mathrm{F}\left(\gamma-\beta\right)^2}{r_\mathrm{FN}},
\end{align}
which is just Eq.~(10) of Rashba's paper.~\cite{Rashba02a} The imaginary part in the expansion of $\mathcal{Z}_\mathrm{n-eq}(\omega)$ can be written as
\begin{equation}\label{xfnrashba} \mathrm{Im}\left(\mathcal{Z}_\mathrm{n-eq}\right)=\omega{R}^2C_\mathrm{diff},
\end{equation}
where the diffusion capacitance is defined as
\begin{equation}\label{cdiff}
\begin{split}
C_\mathrm{diff}=&\frac{1}{2R^2r_\mathrm{FN}^2}
\big\{T_1^\mathrm{N}r_\mathrm{N}\left(\gamma{r}_\mathrm{b}^\ast+\beta{r}_\mathrm{F}\right)^2 \\
&\qquad\quad+T_1^\mathrm{F}r_\mathrm{F}\left[\gamma{r}_\mathrm{b}^\ast
-\beta\left(r_\mathrm{b}^\ast+r_\mathrm{N}\right)\right]^2\big\}
\end{split}
\end{equation}
and $R$ defined as $R=R_\mathrm{n-eq}+(1-\gamma^2)r_\mathrm{b}^\ast$.

%\begin{equation}
%\tilde{Z}^\mathrm{(FN)}=\beta^2\rho_\mathrm{F}^\ast\tilde{L}_\mathrm{F}(\omega)
%+\gamma^2r_\mathrm{b}^\ast-\frac{\left[\beta\rho_\mathrm{F}^\ast\tilde{L}_\mathrm{F}(\omega)+
%\gamma{r}_\mathrm{b}^\ast\right]^2}{\rho_\mathrm{F}^\ast\tilde{L}_\mathrm{F}(\omega)+
%r_\mathrm{b}^\ast+\rho_\mathrm{N}^\ast\tilde{L}_\mathrm{N}(\omega)}
%\end{equation}

\bibliography{SAC_bib}

\providecommand{\noopsort}[1]{}\providecommand{\singleletter}[1]{#1}%
\begin{thebibliography}{28}%
\makeatletter
\providecommand \@ifxundefined [1]{%
 \@ifx{#1\undefined}
}%
\providecommand \@ifnum [1]{%
 \ifnum #1\expandafter \@firstoftwo
 \else \expandafter \@secondoftwo
 \fi
}%
\providecommand \@ifx [1]{%
 \ifx #1\expandafter \@firstoftwo
 \else \expandafter \@secondoftwo
 \fi
}%
\providecommand \natexlab [1]{#1}%
\providecommand \enquote  [1]{``#1''}%
\providecommand \bibnamefont  [1]{#1}%
\providecommand \bibfnamefont [1]{#1}%
\providecommand \citenamefont [1]{#1}%
\providecommand \href@noop [0]{\@secondoftwo}%
\providecommand \href [0]{\begingroup \@sanitize@url \@href}%
\providecommand \@href[1]{\@@startlink{#1}\@@href}%
\providecommand \@@href[1]{\endgroup#1\@@endlink}%
\providecommand \@sanitize@url [0]{\catcode `\\12\catcode `\$12\catcode
  `\&12\catcode `\#12\catcode `\^12\catcode `\_12\catcode `\%12\relax}%
\providecommand \@@startlink[1]{}%
\providecommand \@@endlink[0]{}%
\providecommand \url  [0]{\begingroup\@sanitize@url \@url }%
\providecommand \@url [1]{\endgroup\@href {#1}{\urlprefix }}%
\providecommand \urlprefix  [0]{URL }%
\providecommand \Eprint [0]{\href }%
\providecommand \doibase [0]{http://dx.doi.org/}%
\providecommand \selectlanguage [0]{\@gobble}%
\providecommand \bibinfo  [0]{\@secondoftwo}%
\providecommand \bibfield  [0]{\@secondoftwo}%
\providecommand \translation [1]{[#1]}%
\providecommand \BibitemOpen [0]{}%
\providecommand \bibitemStop [0]{}%
\providecommand \bibitemNoStop [0]{.\EOS\space}%
\providecommand \EOS [0]{\spacefactor3000\relax}%
\providecommand \BibitemShut  [1]{\csname bibitem#1\endcsname}%
\let\auto@bib@innerbib\@empty
%</preamble>
\bibitem [{\citenamefont {Neamen}(2012)}]{Neamen2012}%
  \BibitemOpen
  \bibfield  {author} {\bibinfo {author} {\bibfnamefont {D.~A.}\ \bibnamefont
  {Neamen}},\ }\href@noop {} {\emph {\bibinfo {title} {Semiconductor Physics
  and Devices: Basic Principles}}},\ \bibinfo {edition} {4th}\ ed.\ (\bibinfo
  {publisher} {McGraw-Hill},\ \bibinfo {address} {New York},\ \bibinfo {year}
  {2012})\BibitemShut {NoStop}%
\bibitem [{\citenamefont {Theis}\ and\ \citenamefont
  {Solomon}(2010)}]{Theis2010}%
  \BibitemOpen
  \bibfield  {author} {\bibinfo {author} {\bibfnamefont {T.~N.}\ \bibnamefont
  {Theis}}\ and\ \bibinfo {author} {\bibfnamefont {P.~M.}\ \bibnamefont
  {Solomon}},\ }\bibfield  {title} {\enquote {\bibinfo {title} {In quest of the
  ``next switch'': Prospects for greatly reduced power dissipation in a
  successor to the silicon field-effect transistor},}\ }\href {\doibase
  10.1109/JPROC.2010.2066531} {\bibfield  {journal} {\bibinfo  {journal} {Proc.
  IEEE}\ }\textbf {\bibinfo {volume} {98}},\ \bibinfo {pages} {2005} (\bibinfo
  {year} {2010})}\BibitemShut {NoStop}%
\bibitem [{\citenamefont {Landry}\ \emph {et~al.}(2001)\citenamefont {Landry},
  \citenamefont {Dong}, \citenamefont {Du}, \citenamefont {Xiang},\ and\
  \citenamefont {Xiao}}]{Xiao01}%
  \BibitemOpen
  \bibfield  {author} {\bibinfo {author} {\bibfnamefont {G.}~\bibnamefont
  {Landry}}, \bibinfo {author} {\bibfnamefont {Y.}~\bibnamefont {Dong}},
  \bibinfo {author} {\bibfnamefont {J.}~\bibnamefont {Du}}, \bibinfo {author}
  {\bibfnamefont {X.}~\bibnamefont {Xiang}}, \ and\ \bibinfo {author}
  {\bibfnamefont {J.~Q.}\ \bibnamefont {Xiao}},\ }\bibfield  {title} {\enquote
  {\bibinfo {title} {Interfacial capacitance effects in magnetic tunneling
  junctions},}\ }\href@noop {} {\bibfield  {journal} {\bibinfo  {journal}
  {Appl. Phys. Lett.}\ }\textbf {\bibinfo {volume} {78}},\ \bibinfo {pages}
  {501} (\bibinfo {year} {2001})}\BibitemShut {NoStop}%
\bibitem [{\citenamefont {Chui}\ and\ \citenamefont {Hu}(2002)}]{Chui02}%
  \BibitemOpen
  \bibfield  {author} {\bibinfo {author} {\bibfnamefont {S.~T.}\ \bibnamefont
  {Chui}}\ and\ \bibinfo {author} {\bibfnamefont {L.}~\bibnamefont {Hu}},\
  }\bibfield  {title} {\enquote {\bibinfo {title} {ac transport in
  ferromagnetic tunnel junctions},}\ }\href@noop {} {\bibfield  {journal}
  {\bibinfo  {journal} {Appl. Phys. Lett.}\ }\textbf {\bibinfo {volume} {80}},\
  \bibinfo {pages} {273} (\bibinfo {year} {2002})}\BibitemShut {NoStop}%
\bibitem [{\citenamefont {Kaiju}\ \emph {et~al.}(2002)\citenamefont {Kaiju},
  \citenamefont {Fujita}, \citenamefont {Morozumi},\ and\ \citenamefont
  {Shiiki}}]{Kaiju2002}%
  \BibitemOpen
  \bibfield  {author} {\bibinfo {author} {\bibfnamefont {H.}~\bibnamefont
  {Kaiju}}, \bibinfo {author} {\bibfnamefont {S.}~\bibnamefont {Fujita}},
  \bibinfo {author} {\bibfnamefont {T.}~\bibnamefont {Morozumi}}, \ and\
  \bibinfo {author} {\bibfnamefont {K.}~\bibnamefont {Shiiki}},\ }\bibfield
  {title} {\enquote {\bibinfo {title} {Magnetocapacitance effect of spin
  tunneling junctions},}\ }\href {\doibase 10.1063/1.1451754} {\bibfield
  {journal} {\bibinfo  {journal} {J. Appl. Phys.}\ }\textbf {\bibinfo {volume}
  {91}},\ \bibinfo {pages} {7430} (\bibinfo {year} {2002})}\BibitemShut
  {NoStop}%
\bibitem [{\citenamefont {Chien}\ \emph {et~al.}(2006)\citenamefont {Chien},
  \citenamefont {Lo}, \citenamefont {Hsieh}, \citenamefont {Yao}, \citenamefont
  {Han}, \citenamefont {Zeng}, \citenamefont {Peng},\ and\ \citenamefont
  {Lin}}]{Chien2006}%
  \BibitemOpen
  \bibfield  {author} {\bibinfo {author} {\bibfnamefont {W.~C.}\ \bibnamefont
  {Chien}}, \bibinfo {author} {\bibfnamefont {C.~K.}\ \bibnamefont {Lo}},
  \bibinfo {author} {\bibfnamefont {L.~C.}\ \bibnamefont {Hsieh}}, \bibinfo
  {author} {\bibfnamefont {Y.~D.}\ \bibnamefont {Yao}}, \bibinfo {author}
  {\bibfnamefont {X.~F.}\ \bibnamefont {Han}}, \bibinfo {author} {\bibfnamefont
  {Z.~M.}\ \bibnamefont {Zeng}}, \bibinfo {author} {\bibfnamefont {T.~Y.}\
  \bibnamefont {Peng}}, \ and\ \bibinfo {author} {\bibfnamefont
  {P.}~\bibnamefont {Lin}},\ }\bibfield  {title} {\enquote {\bibinfo {title}
  {Enhancement and inverse behaviors of magnetoimpedance in a magnetotunneling
  junction by driving frequency},}\ }\href {\doibase 10.1063/1.2374807}
  {\bibfield  {journal} {\bibinfo  {journal} {Appl. Phys. Lett.}\ }\textbf
  {\bibinfo {volume} {89}},\ \bibinfo {pages} {202515} (\bibinfo {year}
  {2006})}\BibitemShut {NoStop}%
\bibitem [{\citenamefont {Padhan}\ \emph {et~al.}(2007)\citenamefont {Padhan},
  \citenamefont {LeClair}, \citenamefont {Gupta}, \citenamefont {Tsunekawa},\
  and\ \citenamefont {Djayaprawira}}]{Gupta07}%
  \BibitemOpen
  \bibfield  {author} {\bibinfo {author} {\bibfnamefont {P.}~\bibnamefont
  {Padhan}}, \bibinfo {author} {\bibfnamefont {P.}~\bibnamefont {LeClair}},
  \bibinfo {author} {\bibfnamefont {A.}~\bibnamefont {Gupta}}, \bibinfo
  {author} {\bibfnamefont {K.}~\bibnamefont {Tsunekawa}}, \ and\ \bibinfo
  {author} {\bibfnamefont {D.~D.}\ \bibnamefont {Djayaprawira}},\ }\bibfield
  {title} {\enquote {\bibinfo {title} {Frequency-dependent magnetoresistance
  and magnetocapacitance properties of magnetic tunnel junctions with {M}g{O}
  tunnel barrier},}\ }\href@noop {} {\bibfield  {journal} {\bibinfo  {journal}
  {Appl. Phys. Lett.}\ }\textbf {\bibinfo {volume} {90}},\ \bibinfo {pages}
  {142105} (\bibinfo {year} {2007})}\BibitemShut {NoStop}%
\bibitem [{\citenamefont {Chang}\ \emph {et~al.}(2010)\citenamefont {Chang},
  \citenamefont {Li}, \citenamefont {Huang}, \citenamefont {Tung},
  \citenamefont {Tong},\ and\ \citenamefont {Lin}}]{Chang2010}%
  \BibitemOpen
  \bibfield  {author} {\bibinfo {author} {\bibfnamefont {Y.-M.}\ \bibnamefont
  {Chang}}, \bibinfo {author} {\bibfnamefont {K.-S.}\ \bibnamefont {Li}},
  \bibinfo {author} {\bibfnamefont {H.}~\bibnamefont {Huang}}, \bibinfo
  {author} {\bibfnamefont {M.-J.}\ \bibnamefont {Tung}}, \bibinfo {author}
  {\bibfnamefont {S.-Y.}\ \bibnamefont {Tong}}, \ and\ \bibinfo {author}
  {\bibfnamefont {M.-T.}\ \bibnamefont {Lin}},\ }\bibfield  {title} {\enquote
  {\bibinfo {title} {Extraction of the tunnel magnetocapacitance with
  two-terminal measurements},}\ }\href {\doibase 10.1063/1.3407509} {\bibfield
  {journal} {\bibinfo  {journal} {J. Appl. Phys.}\ }\textbf {\bibinfo {volume}
  {107}},\ \bibinfo {pages} {093904} (\bibinfo {year} {2010})}\BibitemShut
  {NoStop}%
\bibitem [{\citenamefont {Kaiju}\ \emph {et~al.}(2015)\citenamefont {Kaiju},
  \citenamefont {Takei}, \citenamefont {Misawa}, \citenamefont {Nagahama},
  \citenamefont {Nishii},\ and\ \citenamefont {Xiao}}]{Kaiju2015}%
  \BibitemOpen
  \bibfield  {author} {\bibinfo {author} {\bibfnamefont {H.}~\bibnamefont
  {Kaiju}}, \bibinfo {author} {\bibfnamefont {M.}~\bibnamefont {Takei}},
  \bibinfo {author} {\bibfnamefont {T.}~\bibnamefont {Misawa}}, \bibinfo
  {author} {\bibfnamefont {T.}~\bibnamefont {Nagahama}}, \bibinfo {author}
  {\bibfnamefont {J.}~\bibnamefont {Nishii}}, \ and\ \bibinfo {author}
  {\bibfnamefont {G.}~\bibnamefont {Xiao}},\ }\bibfield  {title} {\enquote
  {\bibinfo {title} {Large magnetocapacitance effect in magnetic tunnel
  junctions based on debye-fr\"{o}hlich model},}\ }\href {\doibase
  10.1063/1.4932093} {\bibfield  {journal} {\bibinfo  {journal} {Appl. Phy.
  Lett.}\ }\textbf {\bibinfo {volume} {107}},\ \bibinfo {pages} {132405}
  (\bibinfo {year} {2015})}\BibitemShut {NoStop}%
\bibitem [{\citenamefont {Parui}\ \emph {et~al.}(2016)\citenamefont {Parui},
  \citenamefont {Ribeiro}, \citenamefont {Atxabal}, \citenamefont
  {Bedoya-Pinto}, \citenamefont {Sun}, \citenamefont {Llopis}, \citenamefont
  {Casanova},\ and\ \citenamefont {Hueso}}]{Parui2016}%
  \BibitemOpen
  \bibfield  {author} {\bibinfo {author} {\bibfnamefont {S.}~\bibnamefont
  {Parui}}, \bibinfo {author} {\bibfnamefont {M.}~\bibnamefont {Ribeiro}},
  \bibinfo {author} {\bibfnamefont {A.}~\bibnamefont {Atxabal}}, \bibinfo
  {author} {\bibfnamefont {A.}~\bibnamefont {Bedoya-Pinto}}, \bibinfo {author}
  {\bibfnamefont {X.}~\bibnamefont {Sun}}, \bibinfo {author} {\bibfnamefont
  {R.}~\bibnamefont {Llopis}}, \bibinfo {author} {\bibfnamefont
  {F.}~\bibnamefont {Casanova}}, \ and\ \bibinfo {author} {\bibfnamefont
  {L.~E.}\ \bibnamefont {Hueso}},\ }\bibfield  {title} {\enquote {\bibinfo
  {title} {Frequency driven inversion of tunnel magnetoimpedance and
  observation of positive tunnel magnetocapacitance in magnetic tunnel
  junctions},}\ }\href {\doibase 10.1063/1.4960202} {\bibfield  {journal}
  {\bibinfo  {journal} {Appl. Phys. Lett.}\ }\textbf {\bibinfo {volume}
  {109}},\ \bibinfo {pages} {052401} (\bibinfo {year} {2016})}\BibitemShut
  {NoStop}%
\bibitem [{\citenamefont {Lee}\ and\ \citenamefont {Chen}(2015)}]{Lee15}%
  \BibitemOpen
  \bibfield  {author} {\bibinfo {author} {\bibfnamefont {T.-H.}\ \bibnamefont
  {Lee}}\ and\ \bibinfo {author} {\bibfnamefont {C.-D.}\ \bibnamefont {Chen}},\
  }\bibfield  {title} {\enquote {\bibinfo {title} {Probing spin accumulation
  induced magnetocapacitance in a single electron transistor},}\ }\href
  {\doibase 10.1038/srep13704} {\bibfield  {journal} {\bibinfo  {journal} {Sci.
  Rep.}\ }\textbf {\bibinfo {volume} {5}},\ \bibinfo {pages} {13704} (\bibinfo
  {year} {2015})}\BibitemShut {NoStop}%
\bibitem [{\citenamefont {Rashba}(2002)}]{Rashba02a}%
  \BibitemOpen
  \bibfield  {author} {\bibinfo {author} {\bibfnamefont {E.~I.}\ \bibnamefont
  {Rashba}},\ }\bibfield  {title} {\enquote {\bibinfo {title} {Complex
  impedance of a spin injection junction},}\ }\href@noop {} {\bibfield
  {journal} {\bibinfo  {journal} {Appl. Phys. Lett.}\ }\textbf {\bibinfo
  {volume} {80}},\ \bibinfo {pages} {2329} (\bibinfo {year}
  {2002})}\BibitemShut {NoStop}%
\bibitem [{\citenamefont {Luryi}(1988)}]{Luryi88}%
  \BibitemOpen
  \bibfield  {author} {\bibinfo {author} {\bibfnamefont {S.}~\bibnamefont
  {Luryi}},\ }\bibfield  {title} {\enquote {\bibinfo {title} {Quantum
  capacitance devices},}\ }\href@noop {} {\bibfield  {journal} {\bibinfo
  {journal} {Appl. Phys. Lett.}\ }\textbf {\bibinfo {volume} {52}},\ \bibinfo
  {pages} {501} (\bibinfo {year} {1988})}\BibitemShut {NoStop}%
\bibitem [{\citenamefont {Datta}(2005)}]{Datta2005}%
  \BibitemOpen
  \bibfield  {author} {\bibinfo {author} {\bibfnamefont {S.}~\bibnamefont
  {Datta}},\ }\href@noop {} {\emph {\bibinfo {title} {Quantum transport: atom
  to transistor}}}\ (\bibinfo  {publisher} {Cambridge University Press},\
  \bibinfo {address} {Cambridge, UK},\ \bibinfo {year} {2005})\BibitemShut
  {NoStop}%
\bibitem [{\citenamefont {Kopp}\ and\ \citenamefont
  {Mannhart}(2009)}]{Kopp2009}%
  \BibitemOpen
  \bibfield  {author} {\bibinfo {author} {\bibfnamefont {T.}~\bibnamefont
  {Kopp}}\ and\ \bibinfo {author} {\bibfnamefont {J.}~\bibnamefont
  {Mannhart}},\ }\bibfield  {title} {\enquote {\bibinfo {title} {Calculation of
  the capacitances of conductors: Perspectives for the optimization of
  electronic devices},}\ }\href {\doibase 10.1063/1.3197246} {\bibfield
  {journal} {\bibinfo  {journal} {J. Appl. Phys.}\ }\textbf {\bibinfo {volume}
  {106}},\ \bibinfo {pages} {064504} (\bibinfo {year} {2009})}\BibitemShut
  {NoStop}%
\bibitem [{\citenamefont {Valet}\ and\ \citenamefont {Fert}(1993)}]{vf93}%
  \BibitemOpen
  \bibfield  {author} {\bibinfo {author} {\bibfnamefont {T.}~\bibnamefont
  {Valet}}\ and\ \bibinfo {author} {\bibfnamefont {A.}~\bibnamefont {Fert}},\
  }\bibfield  {title} {\enquote {\bibinfo {title} {Theory of the perpendicular
  magnetoresistance in magnetic multilayers},}\ }\href@noop {} {\bibfield
  {journal} {\bibinfo  {journal} {Phys.\ Rev. B}\ }\textbf {\bibinfo {volume}
  {48}},\ \bibinfo {pages} {7099} (\bibinfo {year} {1993})}\BibitemShut
  {NoStop}%
\bibitem [{\citenamefont {Zhu}, \citenamefont {Hillebrands},\ and\
  \citenamefont {Schneider}(2008)}]{zhu08}%
  \BibitemOpen
  \bibfield  {author} {\bibinfo {author} {\bibfnamefont {Y.~H.}\ \bibnamefont
  {Zhu}}, \bibinfo {author} {\bibfnamefont {B.}~\bibnamefont {Hillebrands}}, \
  and\ \bibinfo {author} {\bibfnamefont {H.~C.}\ \bibnamefont {Schneider}},\
  }\bibfield  {title} {\enquote {\bibinfo {title} {Signal propagation in
  time-dependent spin transport},}\ }\href@noop {} {\bibfield  {journal}
  {\bibinfo  {journal} {Phys.\ Rev. B}\ }\textbf {\bibinfo {volume} {78}},\
  \bibinfo {pages} {054429} (\bibinfo {year} {2008})}\BibitemShut {NoStop}%
\bibitem [{\citenamefont {Zhu}, \citenamefont {Xu},\ and\ \citenamefont
  {Geng}(2014)}]{zhu14}%
  \BibitemOpen
  \bibfield  {author} {\bibinfo {author} {\bibfnamefont {Y.-H.}\ \bibnamefont
  {Zhu}}, \bibinfo {author} {\bibfnamefont {D.-H.}\ \bibnamefont {Xu}}, \ and\
  \bibinfo {author} {\bibfnamefont {A.-C.}\ \bibnamefont {Geng}},\ }\bibfield
  {title} {\enquote {\bibinfo {title} {Charge accumulation due to spin
  transport in magnetic multilayers},}\ }\href@noop {} {\bibfield  {journal}
  {\bibinfo  {journal} {Physica B}\ }\textbf {\bibinfo {volume} {446}},\
  \bibinfo {pages} {43} (\bibinfo {year} {2014})}\BibitemShut {NoStop}%
\bibitem [{\citenamefont {Fert}\ and\ \citenamefont
  {Jaffr\`{e}s}(2001)}]{fert01}%
  \BibitemOpen
  \bibfield  {author} {\bibinfo {author} {\bibfnamefont {A.}~\bibnamefont
  {Fert}}\ and\ \bibinfo {author} {\bibfnamefont {H.}~\bibnamefont
  {Jaffr\`{e}s}},\ }\bibfield  {title} {\enquote {\bibinfo {title} {Conditions
  for efficient spin injection from a ferromagnetic metal to a
  semiconductor},}\ }\href@noop {} {\bibfield  {journal} {\bibinfo  {journal}
  {Phys. Rev. B}\ }\textbf {\bibinfo {volume} {64}},\ \bibinfo {pages} {184420}
  (\bibinfo {year} {2001})}\BibitemShut {NoStop}%
\bibitem [{\citenamefont {Tulapurkar}\ and\ \citenamefont
  {Suzuki}(2011)}]{Tulapurkar11}%
  \BibitemOpen
  \bibfield  {author} {\bibinfo {author} {\bibfnamefont {A.~A.}\ \bibnamefont
  {Tulapurkar}}\ and\ \bibinfo {author} {\bibfnamefont {Y.}~\bibnamefont
  {Suzuki}},\ }\bibfield  {title} {\enquote {\bibinfo {title} {Boltzmann
  approach to dissipation produced by a spin-polarized current},}\ }\href@noop
  {} {\bibfield  {journal} {\bibinfo  {journal} {Phys. Rev. B}\ }\textbf
  {\bibinfo {volume} {83}},\ \bibinfo {pages} {012401} (\bibinfo {year}
  {2011})}\BibitemShut {NoStop}%
\bibitem [{\citenamefont {Juarez-Acosta}\ \emph {et~al.}(2016)\citenamefont
  {Juarez-Acosta}, \citenamefont {Olivares-Robles}, \citenamefont {Bosu},
  \citenamefont {Sakuraba}, \citenamefont {Kubota}, \citenamefont {Takahashi},
  \citenamefont {Takanashi},\ and\ \citenamefont {Bauer}}]{Juarez16}%
  \BibitemOpen
  \bibfield  {author} {\bibinfo {author} {\bibfnamefont {I.}~\bibnamefont
  {Juarez-Acosta}}, \bibinfo {author} {\bibfnamefont {M.~A.}\ \bibnamefont
  {Olivares-Robles}}, \bibinfo {author} {\bibfnamefont {S.}~\bibnamefont
  {Bosu}}, \bibinfo {author} {\bibfnamefont {Y.}~\bibnamefont {Sakuraba}},
  \bibinfo {author} {\bibfnamefont {T.}~\bibnamefont {Kubota}}, \bibinfo
  {author} {\bibfnamefont {S.}~\bibnamefont {Takahashi}}, \bibinfo {author}
  {\bibfnamefont {K.}~\bibnamefont {Takanashi}}, \ and\ \bibinfo {author}
  {\bibfnamefont {G.~E.~W.}\ \bibnamefont {Bauer}},\ }\bibfield  {title}
  {\enquote {\bibinfo {title} {Modelling of the {P}eltier effect in magnetic
  multilayers},}\ }\href {\doibase 10.1063/1.4942163} {\bibfield  {journal}
  {\bibinfo  {journal} {J. Appl. Phys.}\ }\textbf {\bibinfo {volume} {119}},\
  \bibinfo {pages} {073906} (\bibinfo {year} {2016})}\BibitemShut {NoStop}%
\bibitem [{\citenamefont {Wegrowe}\ and\ \citenamefont
  {Drouhin}(2011)}]{Wegrowe2011}%
  \BibitemOpen
  \bibfield  {author} {\bibinfo {author} {\bibfnamefont {J.-E.}\ \bibnamefont
  {Wegrowe}}\ and\ \bibinfo {author} {\bibfnamefont {H.-J.}\ \bibnamefont
  {Drouhin}},\ }\bibfield  {title} {\enquote {\bibinfo {title} {Spin-currents
  and spin-pumping forces for spintronics},}\ }\href {\doibase
  10.3390/e13020316} {\bibfield  {journal} {\bibinfo  {journal} {Entropy}\
  }\textbf {\bibinfo {volume} {13}},\ \bibinfo {pages} {316} (\bibinfo {year}
  {2011})}\BibitemShut {NoStop}%
\bibitem [{\citenamefont {Fert}\ and\ \citenamefont {Lee}(1996)}]{fl96}%
  \BibitemOpen
  \bibfield  {author} {\bibinfo {author} {\bibfnamefont {A.}~\bibnamefont
  {Fert}}\ and\ \bibinfo {author} {\bibfnamefont {S.-F.}\ \bibnamefont {Lee}},\
  }\bibfield  {title} {\enquote {\bibinfo {title} {Theory of the bipolar spin
  switch},}\ }\href@noop {} {\bibfield  {journal} {\bibinfo  {journal} {Phys.
  Rev. B}\ }\textbf {\bibinfo {volume} {53}},\ \bibinfo {pages} {6554}
  (\bibinfo {year} {1996})}\BibitemShut {NoStop}%
\bibitem [{\citenamefont {Zhang}\ \emph
  {et~al.}(2017{\natexlab{a}})\citenamefont {Zhang}, \citenamefont {Zhu},
  \citenamefont {He},\ and\ \citenamefont {Li}}]{Zhang17cpl}%
  \BibitemOpen
  \bibfield  {author} {\bibinfo {author} {\bibfnamefont {X.-X.}\ \bibnamefont
  {Zhang}}, \bibinfo {author} {\bibfnamefont {Y.-H.}\ \bibnamefont {Zhu}},
  \bibinfo {author} {\bibfnamefont {P.-S.}\ \bibnamefont {He}}, \ and\ \bibinfo
  {author} {\bibfnamefont {B.-H.}\ \bibnamefont {Li}},\ }\bibfield  {title}
  {\enquote {\bibinfo {title} {Mechanisms of spin-dependent heat generation in
  spin valves},}\ }\href {\doibase 10.1088/0256-307X/34/6/067202} {\bibfield
  {journal} {\bibinfo  {journal} {Chin. Phys. Lett.}\ }\textbf {\bibinfo
  {volume} {34}},\ \bibinfo {pages} {067202} (\bibinfo {year}
  {2017}{\natexlab{a}})}\BibitemShut {NoStop}%
\bibitem [{\citenamefont {Zhang}\ \emph
  {et~al.}(2017{\natexlab{b}})\citenamefont {Zhang}, \citenamefont {Zhu},
  \citenamefont {He},\ and\ \citenamefont {Li}}]{Zhang17pb}%
  \BibitemOpen
  \bibfield  {author} {\bibinfo {author} {\bibfnamefont {X.-X.}\ \bibnamefont
  {Zhang}}, \bibinfo {author} {\bibfnamefont {Y.-H.}\ \bibnamefont {Zhu}},
  \bibinfo {author} {\bibfnamefont {P.-S.}\ \bibnamefont {He}}, \ and\ \bibinfo
  {author} {\bibfnamefont {B.-H.}\ \bibnamefont {Li}},\ }\bibfield  {title}
  {\enquote {\bibinfo {title} {An alternative to the spin-coupled interface
  resistance for describing heat generation},}\ }\href {\doibase
  10.1016/j.physb.2017.04.004} {\bibfield  {journal} {\bibinfo  {journal}
  {Physica B}\ }\textbf {\bibinfo {volume} {515}},\ \bibinfo {pages} {43}
  (\bibinfo {year} {2017}{\natexlab{b}})}\BibitemShut {NoStop}%
\bibitem [{\citenamefont {Zhang}\ and\ \citenamefont {Levy}(2002)}]{zhang02}%
  \BibitemOpen
  \bibfield  {author} {\bibinfo {author} {\bibfnamefont {S.}~\bibnamefont
  {Zhang}}\ and\ \bibinfo {author} {\bibfnamefont {P.~M.}\ \bibnamefont
  {Levy}},\ }\bibfield  {title} {\enquote {\bibinfo {title} {Time dependence of
  spin accumulation and magnetoresistance in magnetic multilayers},}\
  }\href@noop {} {\bibfield  {journal} {\bibinfo  {journal} {Phys.\ Rev. B}\
  }\textbf {\bibinfo {volume} {65}},\ \bibinfo {pages} {052409} (\bibinfo
  {year} {2002})}\BibitemShut {NoStop}%
\bibitem [{\citenamefont {Zhu}\ and\ \citenamefont {Schneider}(2009)}]{zhu09}%
  \BibitemOpen
  \bibfield  {author} {\bibinfo {author} {\bibfnamefont {Y.~H.}\ \bibnamefont
  {Zhu}}\ and\ \bibinfo {author} {\bibfnamefont {H.~C.}\ \bibnamefont
  {Schneider}},\ }\bibfield  {title} {\enquote {\bibinfo {title} {Theory of
  impedance dynamics for magnetic multilayers},}\ }\href@noop {} {\bibfield
  {journal} {\bibinfo  {journal} {IEEE Trans. Magn.}\ }\textbf {\bibinfo
  {volume} {45}},\ \bibinfo {pages} {3495} (\bibinfo {year}
  {2009})}\BibitemShut {NoStop}%
\bibitem [{\citenamefont {Phan}\ and\ \citenamefont {Peng}(2008)}]{Phan2008}%
  \BibitemOpen
  \bibfield  {author} {\bibinfo {author} {\bibfnamefont {M.-H.}\ \bibnamefont
  {Phan}}\ and\ \bibinfo {author} {\bibfnamefont {H.-X.}\ \bibnamefont
  {Peng}},\ }\bibfield  {title} {\enquote {\bibinfo {title} {Giant
  magnetoimpedance materials: Fundamentals and applications},}\ }\href
  {\doibase 10.1016/j.pmatsci.2007.05.003} {\bibfield  {journal} {\bibinfo
  {journal} {Proc. Mater. Sci.}\ }\textbf {\bibinfo {volume} {53}},\ \bibinfo
  {pages} {323} (\bibinfo {year} {2008})}\BibitemShut {NoStop}%
\end{thebibliography}%

\end{document}